\documentclass[12pt]{article}
\pdfoutput=1
\usepackage{amssymb, amsmath,amsfonts}
\usepackage{graphicx}
\usepackage[linktocpage,pdftex, bookmarks=true,colorlinks,linkcolor=red,urlcolor=blue,citecolor=blue]{hyperref}
\makeatletter\@addtoreset{equation}{section}\makeatother

\def\be{\begin{equation}}
\def\ee{\end{equation}}
\def\bea{\begin{eqnarray}}
\def\eea{\end{eqnarray}}

\newcommand{\nn}{\nonumber}

\def\Dslash{\,\,{\raise.15ex\hbox{/}\mkern-12mu D}}
\def\Dbarslash{\,\,{\raise.15ex\hbox{/}\mkern-12mu {\bar D}}}
\def\delslash{\,\,{\raise.15ex\hbox{/}\mkern-9mu \partial}}
\def\delbarslash{\,\,{\raise.15ex\hbox{/}\mkern-9mu {\bar\partial}}}
\def\pslash{\,\,{\raise.15ex\hbox{/}\mkern-9mu p}}
\def\calDslash{\,\,{\raise.15ex\hbox{/}\mkern-12mu {\cal D}}}

\newcommand{\Tr}{{\rm Tr}}

\makeatletter\@addtoreset{equation}{section}\makeatother

\newcommand{\preprint}[1]{\begin{table}[t]  %%
             \begin{flushright}               %%
             {#1}                             %%
             \end{flushright}                 %%
             \end{table}}                     %%
\renewcommand{\title}[1]{\vbox{\center\LARGE{#1}}\vspace{5mm}}
\renewcommand{\author}[1]{\vbox{\center#1}\vspace{5mm}}
\newcommand{\address}[1]{\vbox{\center\em#1}}

\def\arXiv#1{\href{http://arxiv.org/abs/#1}{arXiv:#1}}
\def\arXiv#1#2{\href{http://arxiv.org/abs/#1}{arXiv:#1}}

\def\doi#1#2{\href{http://doi.org/#1}{#2}}

\begin{document}

\unitlength = .8mm

\begin{titlepage}
\vspace{.5cm}
\preprint{}
 
\begin{center}
\hfill \\
\hfill \\
\vskip 1cm

\title{Topological invariants for holographic semimetals}
\vskip 0.5cm
{Yan Liu$^{a}$}\footnote{Email: {\tt yanliu@buaa.edu.cn}} and
 {Ya-Wen Sun$^{b,c,d}$}\footnote{Email: {\tt yawen.sun@ucas.ac.cn}} 

\address{${}^a$Department of Space Science, and International Research Institute
of Multidisciplinary Science, Beihang University,  Beijing 100191, China}
\address{${}^b$School of physics  \& CAS Center for Excellence in Topological Quantum Computation, University of Chinese Academy of Sciences, Beijing 100049, China}
\address{${}^c$Kavli Insititute for Theoretical Sciences, University of Chinese Academy of Sciences, Beijing 100049,  China}
\address{${}^d$CAS Key Laboratory of Theoretical Physics, Institute of Theoretical Physics, 
\\Chinese Academy of Sciences, Beijing 100190, China}

\end{center}
\vskip 1.5cm

\abstract{We study the behavior of fermion spectral functions for the holographic topological Weyl and nodal line semimetals. We calculate the topological invariants from the Green functions of both holographic semimetals using the topological Hamiltonian method, which calculates topological invariants of strongly interacting systems from an effective Hamiltonian system with the same topological structure. Nontrivial topological invariants for both systems have been obtained and the presence of nontrivial topological invariants further supports the topological nature of the holographic semimetals.}
\vfill

\end{titlepage}

\eject \tableofcontents %

%%%%%%%%%%%%%%%%%%%%%%%%%%%%%%%%%%%%%%%%%
\section{Introduction}
%%%%%%%%%%%%%%%%%%%%%%%%%%%%%%%%%%%%%%%%%
Topological states of matter are a new type of quantum states of matter that cannot be described by the Landau-Ginzburg paradigm and do not possess a local order parameter \cite{Witten:2015aoa}. They are otherwise characterized by nontrivial topological structures in their quantum wave functions and possess novel nontrivial properties that are stable under small perturbations. Many topological states of matter have been found in laboratories already, e.g. topological insulators, anomalous quantum Hall effects, Weyl semimetals, etc.. As most known properties of topological states of matter have been studied in the weakly coupled theory, an important question is if interactions, especially strong interactions, will change the topological properties and destroy the topological structures of these systems. 

In  \cite{Landsteiner:2015lsa, Landsteiner:2015pdh} and \cite{Liu:2018bye},  strongly coupled topological Weyl and nodal line semimetals were found in the framework of anti-de Sitter/conformal field theory (AdS/CFT) correspondence, which turns a strongly coupled field theoretical problem into a weakly coupled classical gravity problem \cite{Zaanen:2015oix, book0,{review}}. The evidence that the holographic Weyl and nodal line semimetals are topological semimetals includes the anomalous Hall conductivity for Weyl semimetals \cite{Landsteiner:2015pdh}, the induced effect of surface state \cite{Ammon:2016mwa}, as well as the nodal loop from the dual fermion spectral functions \cite{Liu:2018bye}. Based on the  holographic models of semimetals, many interesting observations have been made, including a prediction of nontrivial Hall viscosity in the quantum critical region due to the presence of the mixed gauge gravitational anomaly \cite{Landsteiner:2016stv}, the axial anomalous Hall effect \cite{Copetti:2016ewq}, the behavior of AC conductivity \cite{Grignani:2016wyz},  the disorder effect on the topological phase transition \cite{Ammon:2018wzb}, and the properties of quantum chaos in the quantum critical region \cite{Baggioli:2018afg}.\footnote{Different holographic models for Weyl semimetal can be found in \cite{Gursoy:2012ie,Hashimoto:2016ize}.}  Moreover it has been shown that there is a universal bulk topological structure for both holographic topological semimetals \cite{Liu:2018bye}, where the near horizon behavior of the solutions determines that small perturbations could not gap the semimetal phases. However, topological invariants could not be defined associated with the bulk topological structure, and for %the one last piece missing in the evidence 
a further nontrivial piece of evidence 
--- the topological invariants, we have to resort to the dual Green functions obtained from probe fermions on the bulk background. 

For weakly coupled topological systems, topological invariants can be defined from the Bloch states, i.e. the eigenstates of the weakly coupled Hamiltonians. A simple example is the nontrivial Berry phase associated with a closed loop in the momentum space of many topological systems, which is calculated from the Berry connection of the eigenstates of the Hamiltonian. Equivalently, the formula for the topological invariants could also be rewritten using Green functions, which in principle also works at the strong coupling limit. However, the topological invariants defined from Green functions usually require an integral in the imaginary frequency axis, which is extremely time consuming when we only have numerical results for the Green functions. In \cite{{wang-prx}, interaction1, {Wang:2012ig}}, a method called {\em topological Hamiltonian} was developed, which states that topological invariants of a strongly coupled system could be calculated from the eigenstates of an effective Hamiltonian in the same way as in the weakly coupled theory.

As proved in \cite{{wang-prx}, interaction1, {Wang:2012ig}}, this effective topological Hamiltonian could be directly defined from the zero frequency Green functions and it possesses the same topological structure as the original strongly coupled system. Thus to calculate the topological invariants in a strongly coupled holographic semimetal system, we would first need to have the zero frequency Green functions of the fermions that compose the semimetal systems and then calculate the topological invariants from the topological Hamiltonian as if in a weakly coupled system. In this paper we will first probe fermions on the background of the holographic Weyl and nodal line semimetals and calculate the dual retarded Green functions for the fermionic operators, especially focussing on the zero frequency Green functions. Then we obtain the effective topological Hamiltonian and calculate the topological invariants for the holographic semimetals using the topological Hamiltonian method. We will finally show that the holographic semimetals we have obtained indeed possess nontrivial topological invariants.

The paper is organized as follows. We will first review the holographic Weyl and nodal line semimetal models in Sec. \ref{sec2}, which form the basic setups of the topologically nontrivial backgrounds whose topological invariants we will calculate in the paper. In Sec.  \ref{sec3} we will first construct the actions for probe fermions on the background of holographic Weyl and nodal line semimetals separately and then give the prescriptions for calculating retarded Green functions in these two cases. These serve as the basic prescriptions for calculating zero frequency Green functions which we will use In Sec. \ref{sec4} to calculate the topological invariants for both holographic semimetals using the topological Hamiltonian method as the topological Hamiltonian is directly defined from the zero frequency Green functions. Sec. \ref{sec5} is devoted to conclusions and open questions.

%%%%%%%%%%%%%%%%%%%%%%%%%%%%%%%%%%%%%%%%%
\section{Review of holographic Weyl and nodal line semimetals}
\label{sec2}
%%%%%%%%%%%%%%%%%%%%%%%%%%%%%%%%%%%%%%%%%

In this section, we first review the basic setups and the topological structures of the holographic Weyl and nodal line semimetals as well as their phase diagrams. The basics in this section will provide the topologically nontrivial semimetal background for the calculation of topological invariants in Sec \ref{sec4}. More details could be found in \cite{Landsteiner:2015pdh,Liu:2018bye}.

%%%%%%%%%%%%%%%%%%%%%%%%%%%%%
\subsection{Holographic Weyl semimetals}
%%%%%%%%%%%%%%%%%%%%%%%%%%%%%
A Weyl semimetal breaks either time reversal or inversion symmetry \cite{burkov}. For a holographic Weyl semimetal, we have two important fields in the bulk: the axial gauge field $A_{a}$ corresponding to the time reversal symmetry breaking operator whose source intends to separate one Dirac node into two Weyl nodes and a scalar field $\Phi$ corresponding to the Dirac mass operator whose source intends to gap the system. As a mass operator in the field theory breaks the axial symmetry, this scalar field should be axially charged in the bulk with a nonzero source at the boundary that breaks the axial symmetry explicitly. The bulk action of the holographic Weyl semimetal system \cite{Landsteiner:2015pdh} is
\bea
S&=&\int d^5x\sqrt{-g}\bigg[\frac{1}{2\kappa^2}\bigg(R+\frac{12}{L^2}\bigg)-\frac{1}{4}\mathcal{F}^2-\frac{1}{4}F^2+\frac{\alpha}{3}\epsilon^{abcde}A_a \bigg(3\mathcal{F}_{bc}\mathcal{F}_{de}+F_{bc}F_{de}\bigg)
\nn\\
&&~~~-(D_a \Phi)^*(D^a\Phi)-V_1(\Phi)\bigg]\nn\,,
\eea  
where $D_a=\nabla_a -iq_1 A_a$ and $\mathcal{F}_{\mu\nu}$, $F_{\mu\nu}$ are the vector $U(1)_V$ gauge field  strength and  the axial $U(1)_A$ gauge field strength separately. $\alpha$ is the coefficient of the Chern-Simons term which corresponds to the chiral anomaly and $\Phi$ is the axially charge scalar field. The potential term is
\be 
V_1=m_1^2 |\Phi|^2+\frac{\lambda_1}{2} |\Phi|^4\,. 
\ee
We choose the mass of the scalar field to be $m_1^2=-3$ for simplicity.

At zero temperature, the solution can be parametrized as 
\begin{equation}
\label{eq:wsmansatz}
ds^2=u(-dt^2+dx^2+dy^2)+\frac{dr^2}{u}+h dz^2\,,~~~\Phi=\phi\,,~~~ A=A_z dz\,.
\end{equation}
The asymptotic AdS boundary conditions characterizing proper source terms are
\be
\Phi= \frac{M}{r}+\cdots,~~~A_{z}=b+\cdots.
\ee

For general parameter values, there exist three kinds of near horizon solutions at zero temperature, which flow to boundary solutions at three regions of $M/b$. The critical solution corresponds to the near horizon Lifshitz solution, which flows to boundary $M/b=(M/b)_c$. The Weyl semimetal phase has an $AdS_5$ near horizon solution and flows to values of $M/b<(M/b)_c$. The trivial phase has an  $AdS_5$ near horizon solution with a different IR AdS radius and flows to values of $M/b>(M/b)_c$. The different IR AdS radius indicates that some degrees of freedom are gapped out along the RG flow from UV to IR.
Note that $\lambda_1\Phi^4$ at the horizon denotes the degrees of freedom that are not gapped out in the IR.  For the reference of the following sections, we write here the near horizon geometry for the topological phase 
\begin{align}
\label{eq:wsm-nh}
u=r^2,~~
h=r^2,~~
A_z=a_1+\frac{\pi a_1^2\phi_1^2}{16 r} e^{-\frac{2 a_1 }{r}},~~\phi=\sqrt{\pi}\phi_1\Big(\frac{a_1 }{2r}\Big)^{3/2} e^{-\frac{a_1 }{r}}\,,
\end{align} where $a_1$ is the near horizon value of the separation $A_z$ and $\phi_1$ is a free parameter flowing the symmetry to different boundary values of  $M/b<(M/b)_c$.

\begin{figure}[h!]
\begin{center}
\includegraphics[width=0.5\textwidth]{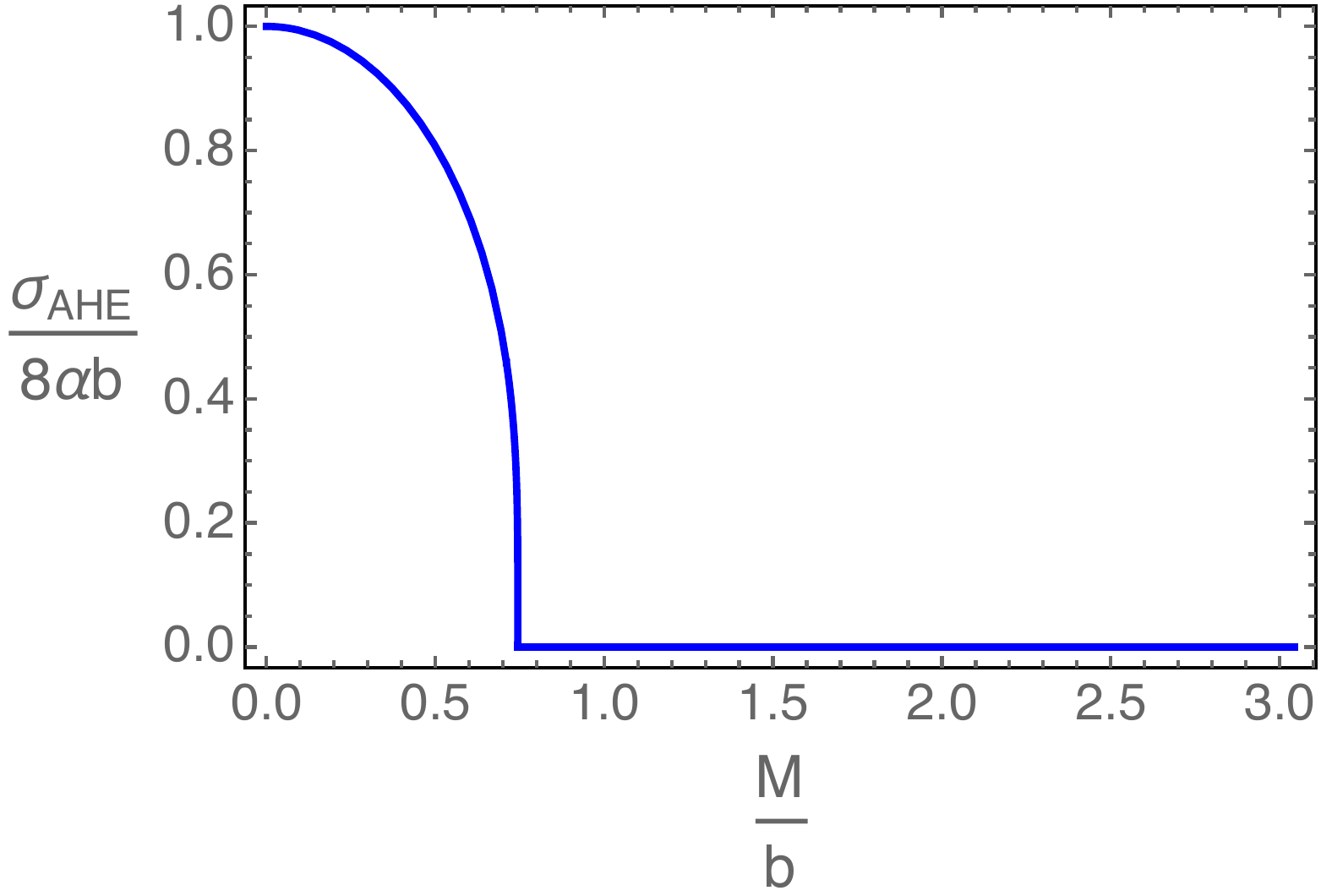}
\end{center}
\vspace{-0.8cm}
\caption{\small The dependance of anomalous Hall conductivity at zero temperature in the holographic Weyl semimetal as a function of $M/b$ for $m_1^2=-3, q_1=1, \lambda_1=1/10$.}
\label{fig:phaseWSM}
\end{figure}
For the Weyl semimetal, there is a smoking gun transport coefficient: the anomalous Hall conductivity $\sigma_\text{AHE}$, which is only nonzero in the Weyl semimetal phase. Semi-analytic calculations showed that $\sigma_\text{AHE}= 8\alpha A_z\big{|}_{r=r_0}$ with the horizon value of $A_z$. Fig \ref{fig:phaseWSM} shows the anomalous Hall conductivity as a function of $M/b$, indicating that the phase at $M/b<(M/b)_c$ is indeed the topological Weyl semimetal phase.

%%%%%%%%%%%%%%%%%%%%%%%%%%%%%
\subsection{Holographic nodal line semimetals}
%%%%%%%%%%%%%%%%%%%%%%%%%%%%%

A nodal line semimetal has a nontrivial shape of Fermi surface where Fermi points connect to form a loop under certain symmetries (see \cite{rev1} for a review). A topologically nontrivial nodal line semimetal cannot be gapped by small perturbations unless passing through a topological phase transition. Two important fields in the holographic setup are the massive two form field $B_{ab}$ whose dual source intends to deform the Dirac point to a nodal loop and the axially charged scalar field whose dual source intends to gap the system.  The action \cite{Liu:2018bye} is 
\bea
S&=&\int d^5x\sqrt{-g}\bigg[\frac{1}{2\kappa^2}\bigg(R+\frac{12}{L^2}\bigg)-\frac{1}{4}\mathcal{F}^2-\frac{1}{4}F^2+\frac{\alpha}{3}\epsilon^{abcde}A_a \bigg(3\mathcal{F}_{bc}\mathcal{F}_{de}+F_{bc}F_{de}\bigg)
\nn\\
&&~~~-(D_a \Phi)^*(D^a\Phi)-V_1(\Phi)-\frac{1}{3\eta}\big(\mathcal{D}_{[a}B_{bc]}\big)^*\big(\mathcal{D}^{[a}B^{bc]}\big)
-V_2(B_{ab})-\lambda|\Phi|^2B_{ab}^*B^{ab}\bigg]\nn
\eea
where $\mathcal{F}_{ab}=\partial_a V_b-\partial_b V_a$ is the vector gauge field strength, $F_{ab}=\partial_a A_b-\partial_b A_a$ is the axial gauge field strength, $D_a=\nabla_a -iq_1 A_a$, $\mathcal{D}_a=\nabla_a -iq_2 A_a$ and
\bea
\mathcal{D}_{[a}B_{bc]}&=&\partial_a B_{bc}+\partial_b B_{ca}+\partial_c B_{ab}-iq_2 A_a B_{bc}-iq_2 A_b B_{ca}-iq_2 A_c B_{ab}\,.
\eea
The potential terms are
\be
V_1=m_1^2 |\Phi|^2+\frac{\lambda_1}{2} |\Phi|^4\,,~~~~
V_2=m_2^2 B^*_{ab}B^{ab}\,,
\ee
where $m_1^2$ is the mass of the scalar field and $m_2^2$ is the mass of the two form field.  The $\lambda$ term denotes the interaction between the scalar field and the two form field. Without loss of generality we choose the conformal dimension for operators dual to $\Phi$ and $B_{ab}$ to be 1, i.e. $m_1^2=-3$ and $m_2^2=1$. We also set $\lambda=1$, $\lambda_1=0.1$ and $\eta=1$ for simplicity. 

Note that here the real part of $\Phi$ corresponds to the operator $\bar{\psi}\psi$ and the imaginary part corresponds to $\bar{\psi}\Gamma^5 \psi$ as could be checked from the ward identity for $J_{\mu}^5$. However, the real and imaginary parts of $B_{ab}$ do not correspond to the composite operators $\bar{\psi}\Gamma_{\mu\nu}\psi$ and $\bar{\psi}\Gamma_{\mu\nu}\Gamma^5\psi$. This is because $\bar{\psi}\Gamma^{\mu\nu}\Gamma^5\psi=\frac{i}{2}\epsilon^{\mu\nu}_{~~\rho\sigma} \bar{\psi}\Gamma^{\rho\sigma}\psi$, which means that the real part and imaginary part of $B_{ab}$ should have a self duality property in order to be dual to $\bar{\psi}\Gamma_{\mu\nu}\psi$ and $\bar{\psi}\Gamma_{\mu\nu}\Gamma^5\psi$.  Here $B_{ab}$ could instead be considered to be dual to a sum of many such kinds of composite operators each composed of a different fermionic operator. In this way, $B_{ab}$ does not need to have the self dual property between its real and imaginary parts. An action that could describe the two form field with the self dual property is $S\propto \int d^5x\sqrt{-g} \big{[} i( B\wedge H^* -B^*\wedge H+m_B^2 |B|^2)\big{]}$, where $H=dB-i q_2 A_5\wedge B$ \cite{Arutyunov:1998xt, Alvares:2011wb}.\footnote{We thank Carlos Hoyos and Elias Kiritsis for helpful discussions on this point.}

The zero temperature solution can be parameterized as 
\be
\label{eq:bg-nlsm}
ds^2 =u(-dt^2+dz^2)+\frac{dr^2}{u}+f(dx^2+dy^2)\,,~~~
\Phi=\phi(r)\,,~~~
B_{xy}=B(r)\,.
\ee
The asymptotic AdS boundary conditions with proper source terms are
\be
\Phi\simeq \frac{M}{r}+\cdots,~~~~B_{xy}\simeq b r+\cdots\,.
\ee

We have three different kinds of near horizon geometries at zero temperature. The critical soluton has a Lifshitz symmetry at the horizon and flows to $M/b=(M/b)_c$ at the boundary. The nodal line semimetal phase has another Lifshitz near horizon solution and flows to values of $M/b<(M/b)_c$. The trivial phase has an $AdS_5$ near horizon solution with a different IR AdS radius and flows to values of $M/b>(M/b)_c$. The different IR AdS radius indicates that some degrees of freedom get gapped out along the RG flow from UV to IR. 
For the reference of following sections, we list the near horzion geometry for the topological phase 
\bea
\label{eq:nh-nlsm}
u&=&\frac{1}{8}(11+3\sqrt{13}) r^2\Big(1+\delta u\, r^{\alpha_1} \Big)\,,\nn\\
f&=& \sqrt{\frac{2\sqrt{13}}{3}-2}\, b_0 r^\alpha \Big(1+\delta f\, r^{\alpha_1} \Big)\,,\nn\\
\phi &=& \phi_0 r^{\beta}\,,\nn\\
B&=&b_0 r^\alpha \Big(1+\delta b\, r^{\alpha_1} \Big)\,,\nn
\eea
where $(\alpha, \beta, \alpha_1)=(0.183, 0.290
, 1.273)$, $(\delta f, \delta b)=(-2.616, -0.302)\delta u$ for the parameter values that we have fixed above.

%Dual fermion spectral function calculations show that there is indeed a nonzero Fermi momentum $k_F=\sqrt{k_x^2+k_y^2}$ in the nodal line semimetal phase. 

%\begin{figure}[h!]
%\begin{center}
%\includegraphics[width=0.5\textwidth]{plot-FS.pdf}
%\end{center}
%\vspace{-0.8cm}
%\caption{\small The dependance of on Fermi surface at zero temperature in the holographic nodal line semimetal semimetal as a function of $M/b$.}
%\label{fig:phaseNLSM}
%\end{figure}

%%%%%%%%%%%%%%%%%%%%%%%%%%%%%
\subsection{A universal bulk topological structure}
%%%%%%%%%%%%%%%%%%%%%%%%%%%%%

There is a universal bulk topological structure for the holographic topological semimetals determined by the horizon solutions. We denote the two kinds of fields as $A$ and $\phi$, the first of which deforms the topology of the Fermi point to whatever possible configurations and the second intends to gap the system. The conformal dimension of the two fields at the horizon are $\delta_{\pm}^{A,\phi}$ separately and the leading order horizon solutions of the two fields are $A,\phi\sim c_{A,\phi} r^{\delta_{+}^{A,\phi}}+\cdots$, where the $r^{\delta_{-}^{A,\phi}}$ terms are too divergent to get regular solutions. The crucial observation is that at the horizon the two coefficients $c_{A,\phi}$ cannot both be nonzero due to the interaction between $A$ and $\phi$ which leads to three different adiabatically connected solutions. %This important feature distinguishes 
The solutions are distingushed into three categories: (1) $c_{A}\neq 0, c_{\phi}=0$; (2) $c_{A}= 0, c_{\phi}\neq 0$, and (3) $c_{A}=0, c_{\phi}=0$, corresponding to three types of phases --- the topological semimetal phase, the partially gapped phase and the critical point. At the horizon $c_{A,\phi}$ cannot coexist leads to the fact that at the semimetal phase, we cannot find a solution of perturbations of the gap operator that could gap the system. Thus small perturbations could not gap the system indicating that the semimetal phases are topological semimetals.

%%%%%%%%%%%%%%%%%%%%%%%%%%%%%%%%%%%%%%%%%%%
\section{Fermion spectral functions of holographic semimetals}
\label{sec3}
%%%%%%%%%%%%%%%%%%%%%%%%%%%%%%%%%%%%%%%%%%%

The existence of a universal bulk topological structure suggests that we could in principle produce a large class of holographic zero density systems which possess a nontrivial topological structure. In some cases we could obtain some specific transport behavior which tells what is the corresponding topological state, e.g. in the Weyl semimetal case, nontrivial anomalous Hall conductivity shows that it corresponds to a topologically nontrivial Weyl semimetal. However, in most cases, we would not be able to tell from the bulk topological structure what would be the boundary topological structure.  In condensed matter physics, the band structure is used to characterize topological structures of weakly coupled topological states of matter. The wave function of electrons or equivalently the Hamiltonian of the system possesses a nontrivial topological structure and topological invariants could be defined. Here for the strongly coupled topological states of matter, there is no band theory or even no quasiparticle descriptions, however, we could still detect the topological structure from the dual Green functions of probe fermions and calculate the topological invariants from the Green functions.\footnote{Fermion spectral function for the holographic finite density systems were first studied in \cite{Liu:2009dm,Cubrovic:2009ye}.} In this section we probe the holographic Weyl/nodal line semimetals with fermions and provide prescriptions for calculating the dual fermion Green functions. The calculations in this section will provide the basic setups for obtaining the topological Hamiltonians for the holographic semimetal systems as the topological Hamiltonian could be directly constructed from the zero frequency Green functions.

%%%%%%%%%%%%%%%%%%%%%%%%%%%%%%%%%%%%%%%%%%%
\subsection{Probe fermions on the holographic Weyl semimetal}
%%%%%%%%%%%%%%%%%%%%%%%%%%%%%%%%%%%%%%%%%%%

To probe the dual fermion spectrum of the holographic Weyl semimetal, we add a probe fermion on the background geometry (\ref{eq:wsmansatz}) and calculate the dual Green functions from the holographic dictionary. In five dimensions, a bulk four component spinor corresponds to a two component chiral spinor of the dual four dimensional field theory \cite{Iqbal:2009fd}. We utilize two spinors $\Psi_1$ and $\Psi_2$ with opposite masses and one standard quantization while the other alternative quantization to correspond to two opposite chiralities.\footnote{Equivalently one could as well choose two spinors with the same mass and the same quantization with the spatial $\Gamma$-matrices of one spinor having an opposite sign compared to the other spinor.}

For the holographic Weyl semimetal, $\Phi$ breaks the axial symmetry so that it couples the left chirality to the right chirality. The axial potential $A_z$ breaks the time reversal symmetry while conserves the axial symmetry, though the two chiralities are affected in different ways by $A_z$. This leads to the following action of probe fermions
\bea\label{eq:probeDirac1}
S&=&S_1+S_2+S_\text{int}\,,\\
S_1&=&\int d^5x \sqrt{-g} i\bar{\Psi}_1\big(\Gamma^a D_a -m_f-i A_a \Gamma^a \big)\Psi_1\,,\nonumber \\
S_2&=&\int d^5x \sqrt{-g} i\bar{\Psi}_2\big(\Gamma^a D_a +m_f +i A_a \Gamma^a \big)\Psi_2\,,\nonumber \\
S_\text{int}&=&-\int d^5x \sqrt{-g}\big( i\eta_1\Phi\bar{\Psi}_1 \Psi_2+i \eta_1^*\Phi^*\bar{\Psi}_2 \Psi_1\big)\,,\nonumber
\eea  
where 
\be
D_a=\partial_a-\frac{i}{4}\omega_{\underline{m}\underline{n},a}\Gamma^{\underline{m}\underline{n}}\,,
\ee 
and we choose both the axial charge and the coupling constant $\eta_1$ to be $1$. Note that the coupling constant in front of $A_z$ is opposite for the two spinors. We use the following convention of $\Gamma$-matrices
\be
\Gamma^{\underline{\mu}}=\gamma^\mu\,,~~~  \Gamma^{\underline{r}}=\gamma^5\,,~~~
\Gamma^{\underline{t}}=\begin{pmatrix}
0 & i \\
i & 0
 \end{pmatrix}\,,~~
 \Gamma^{\underline{i}}=\begin{pmatrix}
0 & i\sigma^i \\
-i\sigma^i & 0
 \end{pmatrix}\,,~~
 \Gamma^{\underline{r}}=\begin{pmatrix}
1 & 0 \\
0 & -1
 \end{pmatrix}\,.~~
\ee

From this form of bulk action for probe fermions, we could see that $\Phi$ corresponds to the operators of $\bar{\psi}\psi$ and $\bar{\psi}\gamma^5\psi$ where $\psi$ is the boundary four component spinor operator. In (\ref{eq:probeDirac1}) $\Phi$ couples to $\bar{\Psi}_1 \Psi_2$, which with $\Psi_{1,2}$ taking opposite quantizations is just the expectation value of the dual operator of $\bar{\psi}\psi$ when the source of $\psi$ is zero. Similar probe fermionic action was considered in \cite{Plantz:2018tqf} to study the holographic mass effect of the four dimensional Dirac fermions.

The equations of motion are 
\bea
\big(\Gamma^a D_a -m_f -i A_z^5 \Gamma^z \big)\Psi_1- \eta_1 \phi \Psi_2=0\,,\nonumber\\
\big(\Gamma^a D_a +m_f+i A_z^5 \Gamma^z \big)\Psi_2-\eta_1 \phi\Psi_1=0\,,
\eea
where we have used $\Phi=\phi(r)$ and $\eta_1$ being a real number. 
We expand the bulk fermion field as
\be
\Psi_l= (uf)^{-1/2} \psi_l e^{-i\omega t+i k_x x+i k_y y+i k_z z} \,,~~~~l=1,2\,.
\ee
%In the background of the Weyl semimetal phase 
Since the spacetime background is isotropic in the $x$-, $y$-plane, after substituting the background geometry the equations of motion for probe fermions become
\be\label{weylf}
\Bigg(\Gamma^{\underline{r}}\partial_r+\frac{1}{u}\Big(-i\omega \Gamma^{\underline{t}}+ik_x\Gamma^{\underline{x}}+ik_y\Gamma^{\underline{y}}\Big)+\frac{1}{\sqrt{uf}}\Big(i(k_z\mp A_z) \Gamma^{\underline{z}}\Big)
+(-1)^l\frac{m_f}{\sqrt{u}}\Bigg)\psi_{l}-\eta_1\frac{\phi}{\sqrt{u}}\psi_{\bar{l}}=0
\ee
with $l=(1,2)$ and $\bar{l}=3-l$. 
 For the Weyl semimetal phase, the equations are isometric in the $x$-$y$ directions and there is a $\omega\to-\omega$ or $k_z\to -k_z$ symmetry.

We can solve (\ref{weylf}) as a set of eight coupled functions. At the horizon the ingoing boundary condition depends on the near horizon geometry. For the topologically trivial phase, the near horizon ingoing solution for nonzero $k$ while $\omega\to 0$ is real just as the pure $AdS_5$ case in \cite{Iqbal:2009fd} and the imaginary part of the Green function is automatically zero where no Fermi surface could be found. For the topologically nontrivial and critical phases, the near horizon ingoing boundary condition is
\be
\psi_l= e^{\frac{i\sqrt{\Delta_l}}{r}}\begin{pmatrix} 
z_1^l  \big(1+\dots\big) \\
\vspace{-.3cm}\\
z_2^l \big(1+\dots\big)\\
\vspace{-.3cm}\\
\frac{i}{\sqrt{\Delta_l}}\big((\omega+k_z+(-1)^l A_z)z_1^l+(k_x-i k_y)z_2^l\big) \big(1+\dots\big)\\
\vspace{-.3cm}\\
\frac{i}{\sqrt{\Delta_l}} \big( (k_x+i k_y)z_1^l+(\omega-(k_z+(-1)^l a_z))z_2^l \big)\big(1+\dots\big) \end{pmatrix}
\ee
with $\Delta_l=\omega^2-k_x^2-k_y^2-(k_z+(-1)^l  a_z)^2$ for  $\omega^2> k_x^2+k_y^2 +(k_z+(-1)^la_z)^2$, 
%\comment{change-CHECK}
%\be
%\psi_{l}\simeq
%\begin{pmatrix} 
%z^{l}_1  \\
%\vspace{-.5cm}\\
%z^{l}_ 2 \\
%\vspace{-.5cm}\\
%i\frac{\sqrt{\omega^2-k_l^2}}{\omega-k_l}z^{l}_1  \\
%\vspace{-.5cm}\\
%i\frac{\sqrt{\omega^2-k_l^2}}{\omega+k_l}z^{l}_2  
%\end{pmatrix}
%  e^{\frac{i\sqrt{w^2-k_l^2}}{u_0r}}\big(1+\cdots\big)\,,~~~~l=(1,2)
%\ee 
%for $\omega>0$, 
where $\cdots$ denotes subleading terms. We will focus on the non-negative frequency and the near horizon boundary condition is only complex when $\omega>k_{1}$ or $\omega>k_2$, where $k_{l}=\sqrt{k_x^2+k_y^2+(k_z+(-1)^l a_0)^2}$ with $a_0$ the horizon value of $A_z$. This is similar to the pure AdS case. 

Near the boundary $r\to\infty$, the Dirac fields behave as 
\be
\psi_{1}=\begin{pmatrix} 
a^{1}_1  &&r^{m_f}+\cdots \\
\vspace{-.5cm}\\
a^{1}_ 2  &&r^{m_f}+\cdots \\
\vspace{-.5cm}\\
a^{1}_ 3 && r^{-m_f}+\cdots \\
\vspace{-.5cm}\\
a^{1}_ 4 && r^{-m_f}+\cdots 
 \end{pmatrix}\,,
 ~~~~\psi_{2}=\begin{pmatrix} 
a^{2}_1  &&r^{-m_f}+\cdots \\
\vspace{-.5cm}\\
a^{2}_ 2  &&r^{-m_f}+\cdots \\
\vspace{-.5cm}\\
a^{2}_ 3 && r^{m_f}+\cdots \\
\vspace{-.5cm}\\
a^{2}_ 4 && r^{m_f}+\cdots  
\end{pmatrix}\,.
\ee Because the two chiralities couple to each other, the source of $\psi_{1,2}$ will also source expectation values of $\psi_{2,1}$. To calculate the retarded Green function, we need four different horizon boundary conditions and get four sets of source and expectation values. We denote the four boundary conditions as I, II, III, IV respectively and the source and expectation matrices are
\vspace{.4cm}\\
$M_s=\begin{pmatrix} 
a^{1,I}_1&~&a^{1,II}_1&~&a^{1,III}_1&~&a^{1,IV}_1\\
\vspace{-.5cm}\\
a^{1,I}_2&~&a^{1,II}_2&~&a^{1,III}_2&~&a^{1,IV}_2\\
\vspace{-.5cm}\\
a^{2,I}_3&~&a^{2,II}_3&~&a^{2,III}_3&~&a^{2,IV}_3\\
\vspace{-.5cm}\\
a^{2,I}_4&~&a^{2,II}_4&~&a^{2,III}_4&~&a^{2,IV}_4\\
\end{pmatrix}$ and $M_e=\begin{pmatrix} 
-a^{2,I}_1&~&-a^{2,II}_1&~&-a^{2,III}_1&~&-a^{2,IV}_1\\
\vspace{-.5cm}\\
-a^{2,I}_2&~&-a^{2,II}_2&~&-a^{2,III}_2&~&-a^{2,IV}_2\\
\vspace{-.5cm}\\
a^{1,I}_3 &~&a^{1,II}_3&~&a^{1,III}_3&~&a^{1,IV}_3\\
\vspace{-.5cm}\\
a^{1,I}_4&~&a^{1,II}_4&~&a^{1,III}_4&~&a^{1,IV}_4\\
\end{pmatrix}\,.$
\vspace{.4cm}\\
The Green function is obtained by $G=i \Gamma ^t M_eM_s^{-1}$. After getting $G$ we find eigenvalues of $G$ and read the imaginary part of the four eigenvalues. We could calculate the retarded Green function using numerics with a very small $\omega$ for numerical convenience.

%%%%%%%%%%%%%%%%%%%%%%%%%%%%%%%%%%%%%%%%%%%
\subsection{Probe fermions on the holographic nodal line semimetal}
%%%%%%%%%%%%%%%%%%%%%%%%%%%%%%%%%%%%%%%%%%%

 The basic setup for the probe fermions on the holographic nodal line semimetal background (\ref{eq:bg-nlsm}) has already been obtained in \cite{Liu:2018bye} and here we will elaborate on more details. The coupling of the two bulk probe spinors to the scalar field is the same as in the Weyl semimetal case while for the holographic nodal line semimetal background, there seem to be multiple consistent ways to couple the two spinors to the $B_{ab}$ field and %we should be careful to choose the 
it turns out that only one way of coupling can deform the Fermi point to a circle. 
%Note that $B_{ab}$ corresponds to the source of the operator $\bar{\psi}\Gamma_{\mu\nu}\psi$, where $\psi$ is the four component boundary spinor. We could write the boundary spinor $\psi$ as $\begin{pmatrix}\psi^L\\ \psi^R\end{pmatrix}$ and $\bar{\psi}\Gamma^{\mu\nu}\psi$ becomes $-\bar{\psi}^{L}\epsilon ^{\mu\nu\rho}\sigma^{\rho}\psi^R$, which means that the composite operator that $B_{ab}$ sources should be of the form $\bar{\psi}^{L}\epsilon ^{\mu\nu\rho}\sigma^{\rho}\psi^R$. 
Expanding $\psi^{L,R}$ to the bulk four component spinor $\psi_{1,2}$, we could write the action of the bulk probe fermions as follows
\bea
S&=&S_1+S_2+S_\text{int}\,,\\
S_1&=&\int d^5x \sqrt{-g} i\bar{\Psi}_1\Big(\Gamma^a D_a -m_f\Big)\Psi_1\,,\nonumber \\
S_2&=&\int d^5x \sqrt{-g} i\bar{\Psi}_2\Big(\Gamma^a D_a +m_f \Big)\Psi_2\,,\nonumber \\
S_\text{int}&=&-\int d^5x \sqrt{-g}\Big( i\Phi\bar{\Psi}_1 \Psi_2+i \Phi^*\bar{\Psi}_2 \Psi_1+
\mathcal{L}_B\Big)\,,
\eea  
and %\comment{covariant form}
\be\label{sb}
\mathcal{L}_B=%i\eta_2 B_{\mu\nu}\bar{\Psi}_1 \epsilon^{\mu\nu\rho}\Gamma_{\rho}C_E\Psi_2-i\eta_{2}^*B_{\mu\nu}^*\bar{\Psi}_2\epsilon^{\mu\nu\rho}\Gamma_{\rho}C_E\Psi_1
-i(\eta_2 B_{ab}\bar{\Psi}_1 \Gamma^{ab}\gamma^5\Psi_2-\eta_{2}^*B_{ab}^*\bar{\Psi}_2\Gamma^{ab}\gamma^5\Psi_1)\,.\ee
%where $C_E=-i\Gamma^t$ and $\mu,\nu,\rho$ indices run over $x,y,z$.  
%The choice of this form of $\mathcal{L}_B$ is to deform the Fermi surface to nodal loops in the holographic NLSM phase. 
Note that the Lorentz invariance in the tangent space has been explicitly broken in the bulk and this is because we have already chosen the source and expectation to correspond to the boundary values of $\Gamma^{\underline{r}} \psi_{1}^{s,e}=\pm \psi_{1}^{s,e}$ and $\Gamma^{\underline{r}} \psi_{2}^{s,e}=\mp \psi_{2}^{s,e}$. %We have explained above why the special choice of $\mathcal{L}_B$ is used and 
Here $\Gamma^{\underline{xy}}\gamma^5$ exchanges the position of the source and expectation spinors of $\Psi_{1,2}$ so that $B_{ab}$ couples to the expectation values of both bulk spinors $\Psi_{1,2}$. If we take two spinors of the same mass and the same quantization, we would not need the $\gamma^5$ matrix in the $\mathcal{L}_B$ term but in the $\Phi$ term to couple the fields to the expectation values of $\Psi^{L,R}$ at the boundary.

There are other physically consistent ways to construct the action of $S_B$, e.g. some possibilities are
\bea
S_{B_1}&=&-\int d^5x \sqrt{-g}(i\eta_2B_{ab}\bar{\Psi}_1 \Gamma^{ab}\Psi_1+i\eta_2B_{ab}^*\bar{\Psi}_2 \Gamma^{ab}\Psi_2)\,,\\
S_{B_2}&=&-\int d^5x \sqrt{-g}(i\eta_2B_{ab}\bar{\Psi}_1 \Gamma^{ab}\Psi_1-i\eta_2B_{ab}^*\bar{\Psi}_2 \Gamma^{ab}\Psi_2)\,,\\
S_{B_3}&=&-\int d^5x \sqrt{-g}(i\eta_2B_{ab}\bar{\Psi}_1 \Gamma^{ab}\Psi_2+i\eta_2B_{ab}^*\bar{\Psi}_2 \Gamma^{ab}\Psi_1)\,.
\eea   
However, all these could not probe the fermion spectral functions of the nodal line semimetal states but other systems where $B_{ab}$ corresponds to the source of other types of composite operators, and only the choice of $\mathcal{L}_B$ in (\ref{sb}) corresponds to a topological nodal line semimetal.

The corresponding Dirac equation can be written as 
\bea
&&\Bigg(\Gamma^{\underline{r}}\partial_r+\frac{1}{u}\Big(-i\omega \Gamma^{\underline{t}}+ik_z \Gamma^{\underline{z}}\Big)+
\frac{1}{\sqrt{uf}}\Big(ik_x\Gamma^{\underline{x}}+ik_y\Gamma^{\underline{y}}\Big)
+(-1)^l\frac{m_f}{\sqrt{u}}\Bigg)\psi_{l}\nn\\
&&~~~~-\Bigg(\eta_1\frac{\Phi}{\sqrt{u}}+(-1)^l\eta_2\frac{b}{\sqrt{u}f}\Gamma^{\underline{x}\underline{y}}\gamma^5\Bigg)\psi_{\bar{l}}=0\, ,
\eea with $l=(1,2)$ and $\bar{l}=3-l$.

The system has an $SO(2)$ symmetry in the $k_x$-$k_y$ plane and only depends on $k_{x-y}=\sqrt{k_x^2+k_y^2}$. Thus without loss of generality we could work at $k_y=0$ in the following. For $k_z\neq 0$ or $\omega\neq 0$ the $k_z$ or $\omega$ terms are more important at the horizon, thus the infalling near horizon boundary conditions are determined by $k_z$ or $\omega$. For $k_z=\omega=0$, the near horizon boundary conditions are determined by the $k_x$ and $k_y$ terms. Then we could obtain the Green functions using the same formula as for the holographic Weyl semimetal phase. 

We could work at $k_z=0$ while $\omega\to 0$ to see the imaginary part of poles in the retarded Green functions. At $\omega=0$ the imaginary part would disappear and the retarded Green functions become real. However, here for the purpose of calculating the topological invariants and also because we could still detect the imaginary poles at $k_z=\omega=0$ which becomes divergences in the real part, we would focus on the $k_z=\omega=0$ data directly. For these poles, when we introduce a very small $\omega$ high peaks of imaginary parts would show up.

\begin{figure}[h!]
\begin{center}
\includegraphics[width=0.63\textwidth]{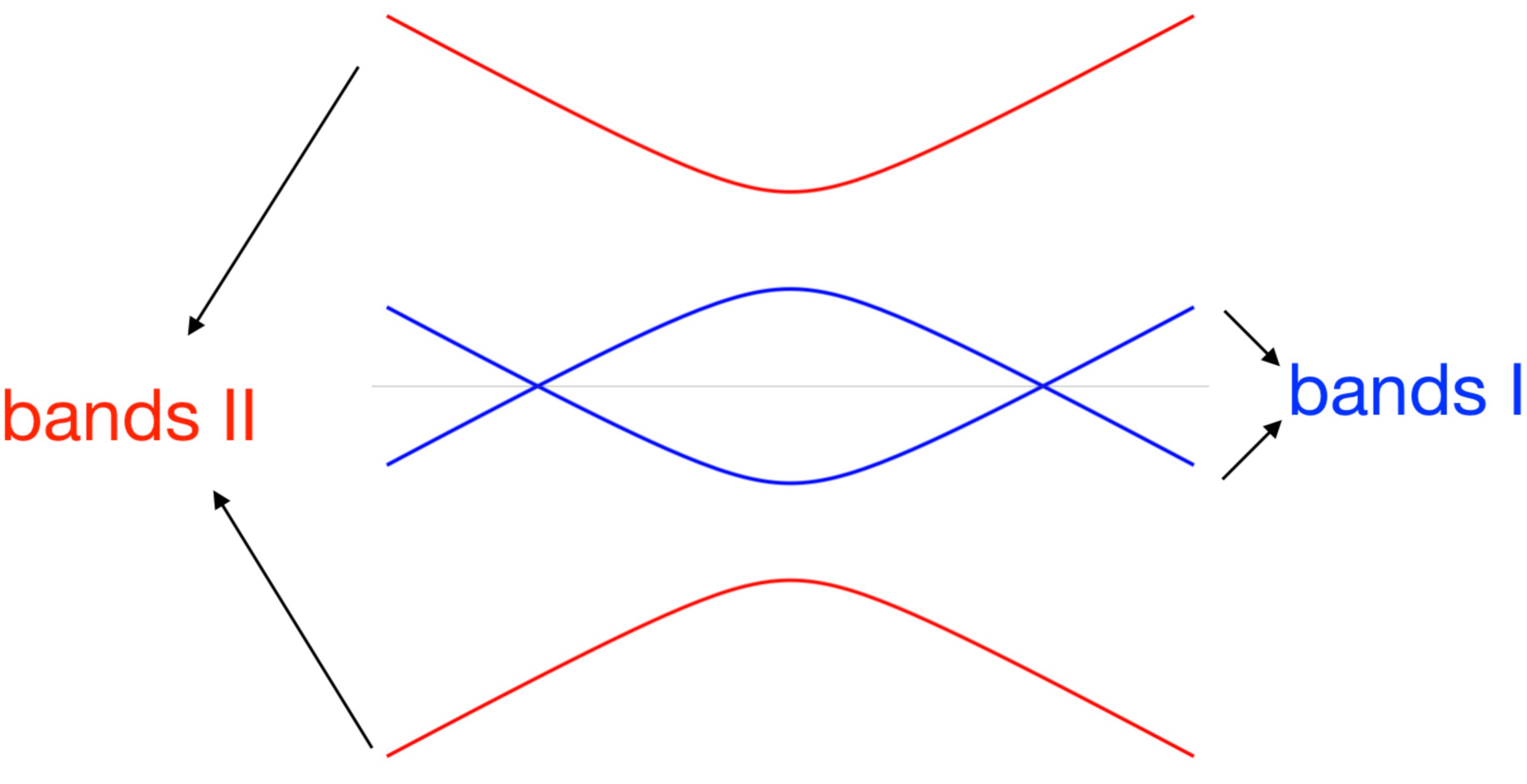}
\end{center}
\vspace{-0.4cm}
\caption{\small Illustration of ``bands" I and II near a $k_{F,i}$ in the $\omega$-$k_x$ plane for $k_y=k_z=0$. The pole $k_{F,i}$ is always a band crossing point of two ``bands".}
\label{fig:cartoonband}
\end{figure}

At zero frequency and $k_z=0$, the four eigenvalues of the Green function are all real and appear in pairs in the form of $(g_1,-g_1,g_2,-g_2)$, where $g_1$ and $g_2$ are positive values and without loss of generality we choose $g_1\geq g_2$. We denote the two branches of eigenstates with eigenvalues $g_1,-g_1$ as ``bands I" and the two branches of eigenstates with eigenvalues $g_2,-g_2$ as ``bands II". An illustration of the four bands in the $\omega$-$k_x$ plane at $k_y=k_z=0$ is in Fig. \ref{fig:cartoonband}. Bands crossings arise when $g_1=g_2$ where bands I and bands II cross at two symmetric points or when $g_1=\infty$ where bands I cross at a pole. 

From numerics we could tell that for background solutions in the nodal line semimetal phase, there are multiple and discrete Fermi nodal lines at $k_{F,i}=\sqrt{k_x^2+k_y^2}$ in the fermion spectral functions at which a pole exists at $\omega=k_z=0$.  At the critical point, $k_F=0$ for $\omega=0$. The nodal lines at $k_{F,i}$ and $\omega=0$ are all band crossing lines of two bands.  At the nodal lines the zero frequency Green functions have two infinite eigenvalues corresponding to these two crossing bands and two other finite and opposite to each other eigenvalues corresponding to the two gapped bands.

\begin{figure}[h!]
\begin{center}
\includegraphics[width=0.4\textwidth]{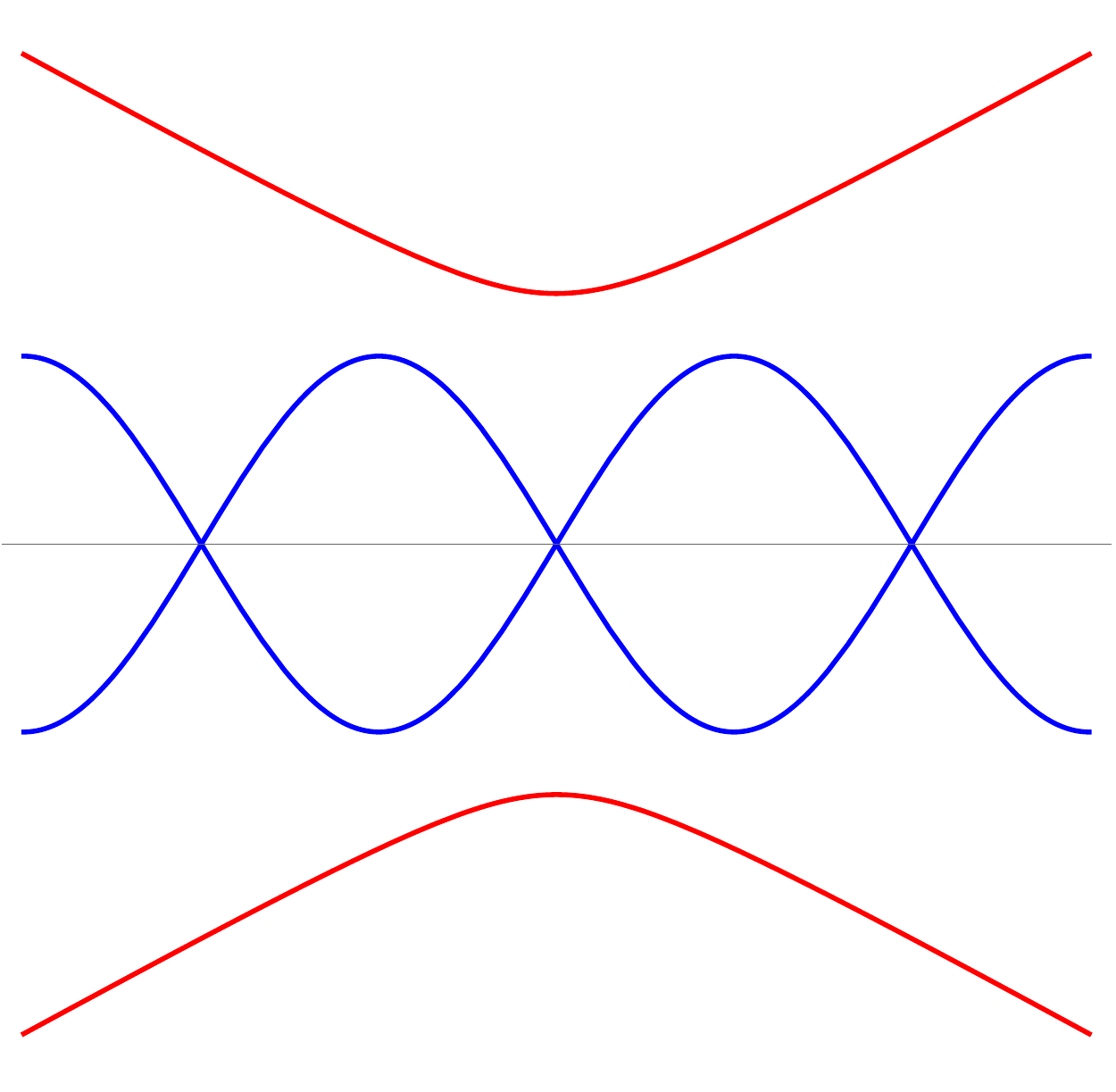}
\includegraphics[width=0.4\textwidth]{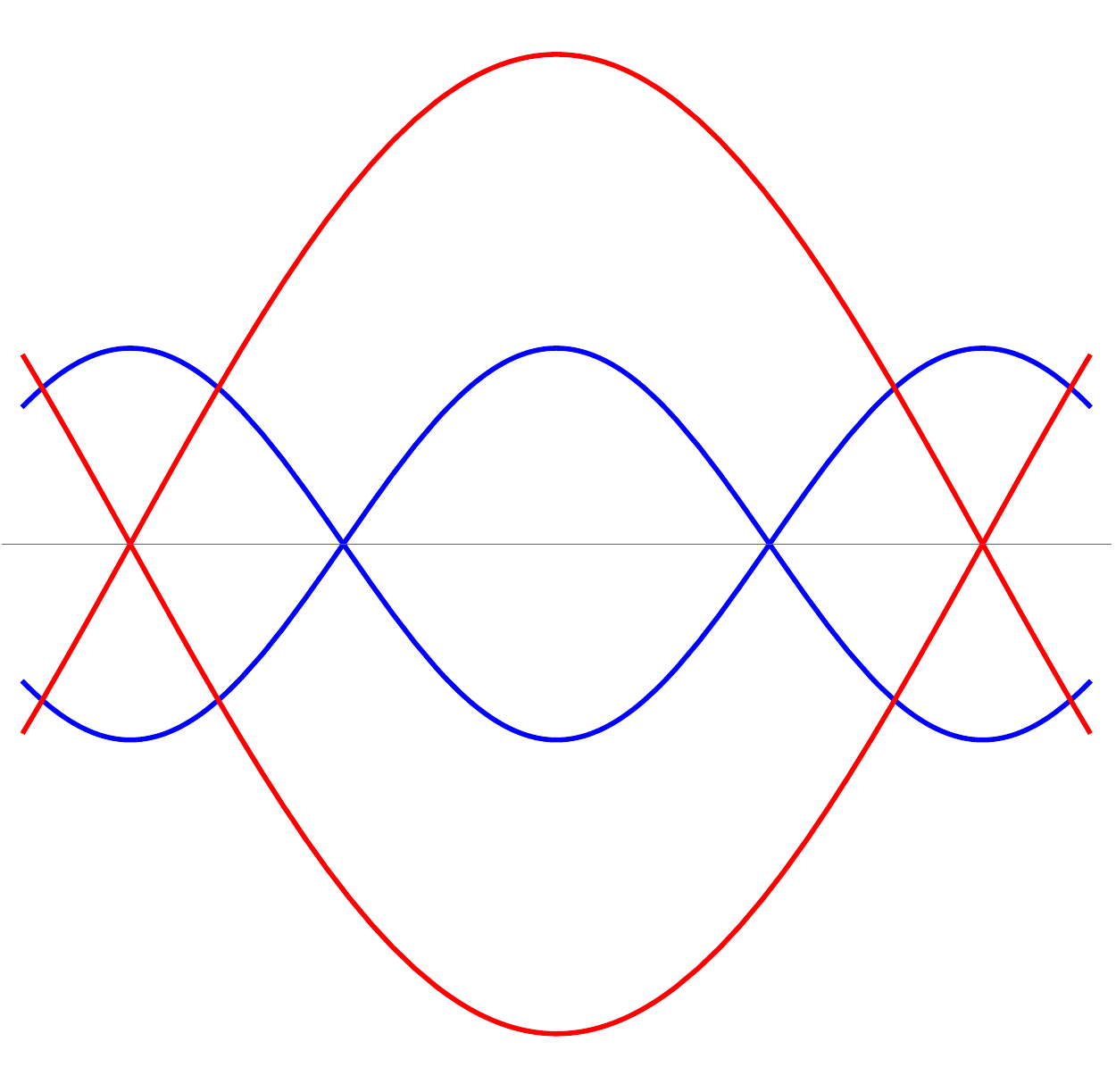}
\end{center}
\vspace{-0.4cm}
\caption{\small Illustration for a multiple-nodal line system in the $E-k_x$ plane where all the poles are from the same two bands (left) or come from different sets of two bands (right).}
\label{fig:cartoon}
\end{figure}

 One immediate question is if these poles all come from the same two bands or different sets of two bands. For the second possibility, gapped bands at a certain $k_{F,i}$ might become gapless poles at another $k_{F,j\neq i}$ and for this to happen, the two sets of bands have to intersect at some points in the $\omega$-$k_x$ plane at $k_y=k_z=0$. Fig. \ref{fig:cartoon} shows the illustration for the spectrum in the $\omega$-$k_x$ plane for a multiple-nodal line system where all the poles are from the same two bands (left) or from different sets of two bands (right).  
%\comment{insert two figures together. Caption: Illustration for a multiple-nodal line system where all the poles are from the same two bands (left) or come from different sets of two bands (right).}

To answer this question, it seems that we would need a spectral density plot of $G(\omega,k)$ for the $\omega$-$k_x$ plane to see which of the following possibility happens: (1) the two sets of bands I and II would intersect at some points in the $\omega$-$k_x$ plane and some of the poles are from bands I and others are from bands II; or (2) the two sets of bands do not intersect for the whole $\omega$-$k_x$ plane and the poles are always from bands I. However, in fact we could distinguish these two possibilities just from the data of the zero frequency Green functions. The explanation is the following. When there is a pole in the zero frequency Green function, i.e. at least one of the eigenvalues reaches infinity, a Fermi point would appear at $\omega=0$ in the spectral density plot for spectral weight of fermions in the $\omega$-$k_x$ plane. The value of the zero frequency Green function eigenvalues reflects how far the band peaks are from the $k_x$ axis in the spectral density plot. When the eigenvalue is small (large), the bands are far away from (close to) the $k_x$ axis. Thus we could use the eigenvalues of $G^{-1}(0,k_x)$ to denote the relative distance of the bands to the $k_x$ axis and plot a qualitative picture of spectral density plot in the $\omega$-$k_x$ plane. In this way, to tell if all the poles come from the same bands or different bands we only need to examine if there is a band crossing point between two adjacent poles at which $g_1=g_2$. If $g_1$ is always larger than $g_2$ when the system evoles from one pole $k_{F,i}$ to the next one $k_{F,i+1}$, then we could tell that the poles are always from bands I, however, if there is a certain $k_{F,i}<k_x<k_{F,i+1}$ at which $g_1=g_2$, the two poles should come from different sets of bands.\footnote{The band crossing points always exist when we tune the value of $M/b$ in the nodal lines semimetal phase, thus the band crossing should not be accidental.} 

%When the absolute value of the four eigenvalues of $G(0,k_x)$ become the same at certain $k_x$, i.e. when $g_1=g_2$ the two bands with positive (negative) eigenvalues cross at  $\omega=\omega_0$ ($\omega=-\omega_0$), $k=k_x$ in the $\omega-k_x$ plane and this is consistent with our illustration to identify $1/G$ as the distance to the $k_x$ axis. 

\begin{figure}[h!]
\begin{center}
\includegraphics[width=0.6\textwidth]{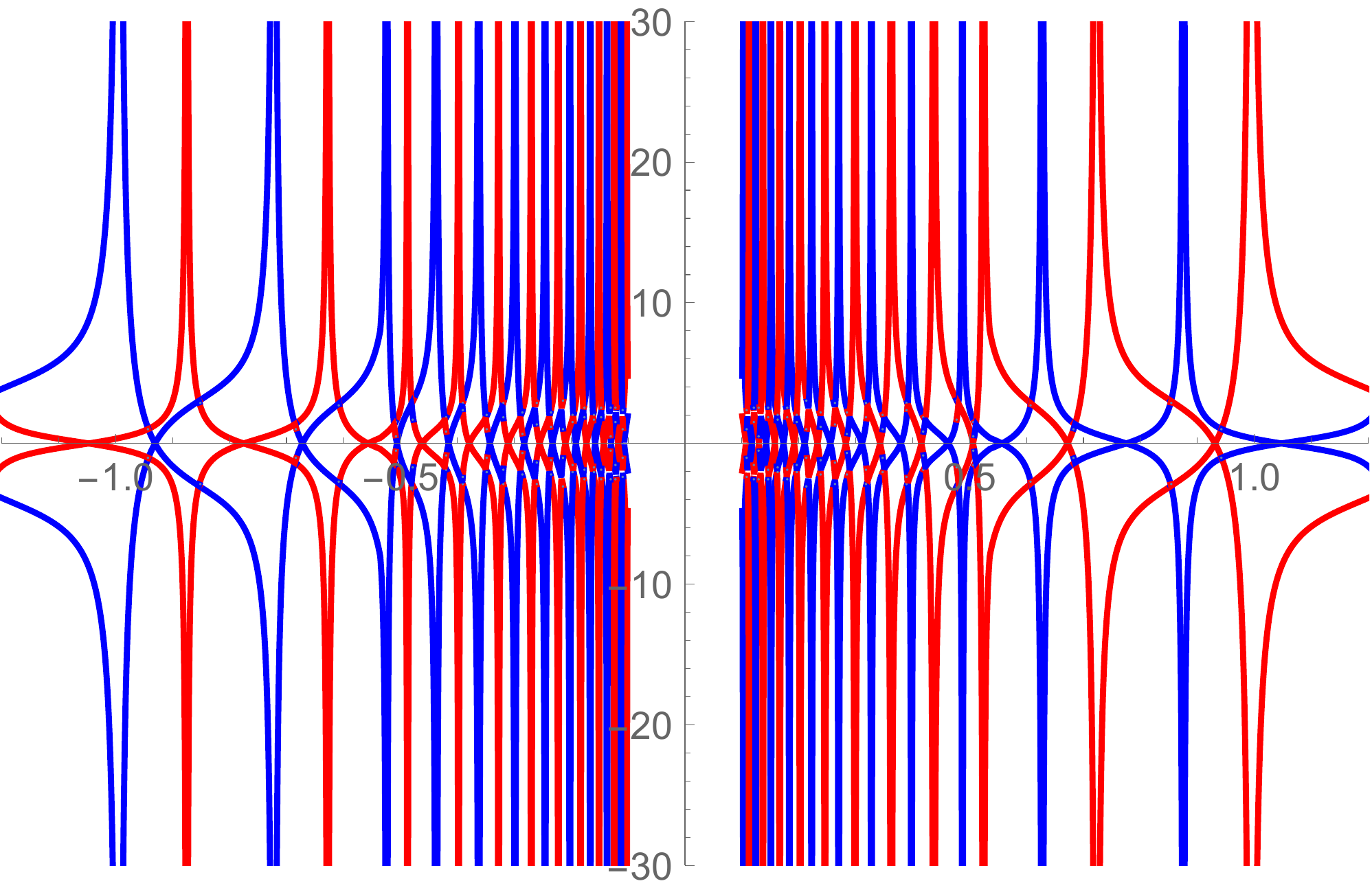}
\end{center}
\vspace{-0.4cm}
\caption{\small Eigenvalues of $-G^{-1}(0,k_x)$ for $M/b\simeq 0.0013$ representing the qualitative behavior of the bands, which should agree qualitatively with the spectral density plot in the $\omega$-$k_x$ plane. We refer to the two bands with red colour as bands I and the two bands with blue colour as bands II. The distance between adjacent poles are becoming larger as $k_x$ increases.}
\label{fig:spec}
\end{figure}

Fig. \ref{fig:spec} is the qualitative behavior of the bands for $M/b\simeq 0.0013$ and $m_f=-1/4$, which should agree qualitatively with the spectral density plot in the $\omega$-$k_x$ plane. We will see in the next section that this is in fact the spectrum/band structure of the topological Hamiltonian defined from the zero frequency Green functions and this is consistent with the main spirit of the topological Hamiltonian approach that the zero frequency Green functions capture all the topological information of the system. 

The first observation is that the distance between adjacent poles are becoming larger as $k_x$ increases. At small $k_x$ the poles are very sharp and very close to each other and we did not plot this area as the nodal loops are so dense that we need to run at extremely small intervals of $k_x$ to reveal all the poles which requires a much larger accuracy.

We could see from the figures that bands I and II always intersect once and only once in the upper $\omega$ plane between each two adjacent poles, which means that the adjacent two poles always come from different two sets of bands. Different from the weakly coupled nodal line semimetal system where the four bands are divided into two gapless bands and two gapped bands which are always gapped, now the two gapped bands in the nodal line semimetal phase are not always gapped but soon become gapless at a larger $k_x$ and exchange the role with the other two bands. Between each adjacent two poles, there is one and only one band crossing point in the upper $\omega$ plane. Another interesting observation is that between each two adjacent band crossing points there is always one pole and one zero of the Green function. This means that for positive $m_f$ there will also be poles. However, we will show below that different from the holographic Weyl semimetal case, these zeros do not possess nontrivial Berry phases. 

\begin{figure}[h!]
\begin{center}
\includegraphics[width=0.5\textwidth]{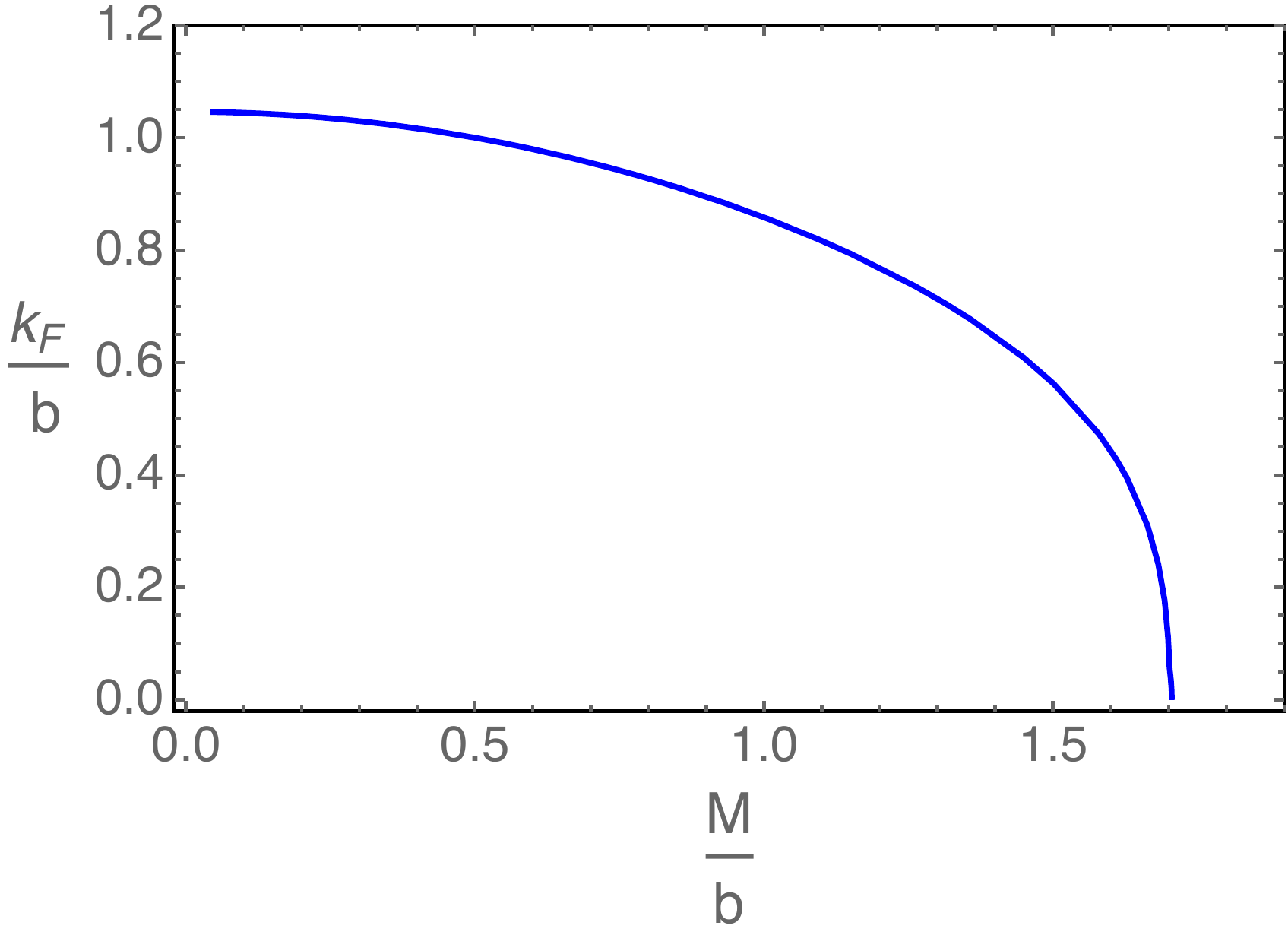}
\end{center}
\vspace{-0.8cm}
\caption{\small The dependance of one branch of the nodal loop radius at zero temperature in the holographic nodal line semimetal phase as a function of $M/b$. Clearly the radius of the nodal loop reaches zero when $M/b$ approaches the critical value. The qualitative behavior is the same for other branches of nodal loops.}
\label{fig:phaseNLSM}
\end{figure}

When we increase $M/b$ to be approaching the critical value of $(M/b)_c$, all the nodal loops would shrink in size and finally become a point at the critical point. Fig. \ref{fig:phaseNLSM} shows the evolution of one $k_F$ as a function of $M/b$. For each of the nodal lines, we have sharp Fermi surface and a linear dispersion in all the $k_x$, $k_y$ and $k_z$ directions \cite{Liu:2018bye}.

%%%%%%%%%%%%%%%%%%%%%%%%%%%%%%%%%%%%%%%%%%%
\section{Topological invariants}
\label{sec4}
%%%%%%%%%%%%%%%%%%%%%%%%%%%%%%%%%%%%%%%%%%%

In mathematics, topological objects possess properties that are invariant under homeomorphisms, which are called topological invariants. Topological invariants could be numbers, e.g. the genus of a closed surface, or could also be groups, e.g. the fundamental group. In the same way, topological invariants could be defined for topological states of matter, which are invariant under adiabatic deformations that do not change the topology of the underlying physical system.

For weakly coupled topological systems, a simple example of a topological invariant is the Berry phase with value $0$ or $\pi$, which is the phase accumulated along a closed loop $\gamma$ in the momentum space for the Bloch states, i.e. eigenstates of the Hamiltonian $|n_{\bf k}\rangle$. The formula for Berry phase \cite{berry} is
\be\label{eq:berryphase}
\phi=\oint_\gamma \mathcal{A}_{\bf k}\cdot d {\bf k}\,,
\ee 
where the Berry connection is defined by eigenstates 
$|n_{\bf k}\rangle$ 
\be 
\mathcal{A}_{\bf k}=i \sum_{j}\langle n_{\bf k}|\partial_{\bf k}|n_{\bf k}\rangle\,, 
\ee  where $j$ runs over all occupied bands and $|n_{\bf k}\rangle$ is the eigenvector of the momentum space Hamiltonian. 
Berry phase could be defined in general dimensions and here we focus on $3+1$ dimensions for our purpose. 
We can also write (\ref{eq:berryphase}) using the Berry curvature as
\be
\phi=\int_S {\bf  \Omega}\cdot d{\bf S}\,,
\ee where 
\be \label{eq:berrycur}
\Omega_i=\epsilon_{ijl } \big( \partial_{k_j}\mathcal{A}_{k_l}-\partial_{k_l}\mathcal{A}_{k_j}\big)
\ee
 and $d{\bf S}$ is the surface element of $S$ which is a surface surrounded by the closed loop $\gamma$, i.e. $\gamma=\partial S$.  
 
An equivalent calculation of this topological invariant is to use the Green function \be
N(k_z)=\frac{1}{24\pi^2}\int dk_0 dk_x d k_y \Tr\Big[\epsilon^{\mu\nu\rho z}G\partial_{\mu}G^{-1}G\partial_{\nu}G^{-1}G\partial_{\rho}G^{-1}\Big]\,,
\ee
where $\mu,\nu,\rho \in k_0,k_x,k_y$ and $k_0=i\omega$ is the Matsubara frequency. For noninteracting systems, the Green function $G(i\omega, k)=1/(i\omega-h(k))$ where $h(k)$ is the Hamiltonian matrix $H=\sum_k c_k^\dagger h(k)c_k$. This formula for the topological invariant is still applicable for interacting systems, however, it involves an integration in the $i\omega$ direction, which is difficult to get in practical strongly coupled systems. This is not a problem in holography as in principle we could get the Green function for any value of $\omega$ using numerics, which, however, is extremely time consuming.

%For both gapped and gapless topological band systems in 3+1D, the nontrivial topology of the quantum wave function in the three dimensional momentum space could be characterized by nontrivial topological invariants. 
%For a closed surface in the momentum space, a similar topological invariant could be defined using the Berry curvature as 
%\comment{CHECK}
%\be 
%c=\frac{1}{2\pi}\oint_S {\bf\mathcal{ F}}\cdot d{\bf S}\,,
%\ee 
%\be 
%C=\frac{1}{2\pi}\oint_S {\bf\Omega}\cdot d{\bf S}\,,
%\ee 
%which is quantized to integer numbers. 

%The topological invariant is $C=\int dk_z N(k_z)$ \cite{interaction1}

In \cite{Wang:2012ig,wang-prx} it was shown that the zero frequency Green function $G(0,{\bf k})$ already contains all the topological information. One could define an effective topological Hamiltonian 
\be \label{eq:topH}
\mathcal{H}_t({\bf k})=-G^{-1}(0,{\bf k})
\ee
and define eigenvectors using this effective topological Hamiltonian. As long as $G(i\omega,{\bf k})$ does not have a pole at nonzero $\omega$, the topological invariants defined under the effective Hamiltonian $\mathcal{H}_t({\bf k})$ as if the system is a weakly coupled theory with the Hamiltonian $\mathcal{H}_t({\bf k})$ would be the same as those defined in the original system. Thus we could define topological invariants using negative valued eigenvectors of $\mathcal{H}_t({\bf k})$, i.e. effective occupied states $n_{\bf k}$ with $\mathcal{H}_t({\bf k})|n_{\bf k}\rangle=-E_t |n_{\bf k}\rangle$ and $E_t>0$.

With the setup in the previous section, we could calculate the zero frequency Green functions for probe fermions and obtain the topological invariants from the Green functions using the method above. When there is no pole in the imaginary $\omega$ axis in the Green function, the topological invariant could be calculated from the weakly coupled formula defined for the effective topological Hamiltonian. Once we have obtained the topological Hamiltonian, the procedure would be the same as the weakly coupled case. 

In the following, we will first obtain the Green function at zero frequency for both the holographic Weyl and nodal line semimetal states and calculate the topological invariants from occupied eigenvectors of the zero frequency Green functions. To understand this procedure easier, we will first start with a simple example, which is the calculation of holographic topological invariants for the pure AdS case before going to the Weyl and nodal line cases.

%%%%%%%%%%%%%%%%%%%%%%%%%%%%%%%%%%%%%%%%%%%
\subsection{Topological invariant for the pure AdS case}
%%%%%%%%%%%%%%%%%%%%%%%%%%%%%%%%%%%%%%%%%%%

In the pure AdS case, the system is in fact degenerate at zero frequency, which is easy to understand as the two Weyl nodes coincide to form a Dirac node, but we could still distinguish the two degenerate eigenstates according to their chiralities. The retarded Green functions for one chirality in the pure AdS case for $\omega>k$ has already been obtained in \cite{Iqbal:2009fd}. For pure AdS, the two chiralities do not interact and we could directly get the full Green function using two spinors of opposite masses and quantizations. In this case, the action of the two spinors are
\bea
S&=&S_1+S_2\,,\\
S_1&=&\int d^5x \sqrt{-g} i\bar{\Psi}_1\big(\Gamma^a D_a -m_f\big)\Psi_1\,,\nonumber \\
S_2&=&\int d^5x \sqrt{-g} i\bar{\Psi}_2\big(\Gamma^a D_a +m_f \big)\Psi_2\,.
\eea
To obtain the topological Hamiltonian, we focus on the $\omega=0$ solutions and the zero frequency Green function. We parametrize the solution as $\Psi_l=(\psi_l^+, \psi_l^-)^T$ with $l=(1,2)$. 
Different from the $\omega>k$ case, at zero frequency, the solutions as well as the Green functions are real functions of $k=\sqrt{k_x^2+k_y^2+k_z^2}$. The solution of this action with infalling boundary condition at the horizon is 
%\be
%\psi^+_{1,2}=r^{-1/2}K_{\pm m_f+\frac{1}{2}}\Big(\frac{k}{r}\Big)a^+_{1,2}\,,
%\ee  
\be
\psi^+_{l}=r^{-1/2}K_{-(-1)^l m_f+\frac{1}{2}}\Big(\frac{k}{r}\Big)a^+_{l}\,,~~~l=(1,2)
\ee 
where $a^+_{1,2}$ are two arbitrary spinors and $K_{\pm m_f+\frac{1}{2}}\Big(\frac{k}{r}\Big)$ is the BesselK function. $\psi^-_{1,2}$ could be obtained %from $\psi^+_{1,2}$ 
from the equations of motion for $\psi^+_{1,2}$, which in our convention of $\Gamma$-matrices is 
%\be
%\psi^-_{1,2}=\frac{k_\mu\sigma^\mu}{k^2}r\big(r\partial_r\mp m_f\big)\psi^+_{1,2}\,.
%\ee
\be
\psi^-_{l}=\frac{k_\mu\sigma^\mu}{k^2}r\big(r\partial_r+(-1)^l m_f\big)\psi^+_{l}\,.
\ee
Four boundary conditions could be identified as four linearly independent choices of $a^+_{1,2}$. After expanding the solutions at the boundary we could get the two source and expectation matrices. The final result for the retarded Green function of two chiralities are
\be\label{eq:specads}
G(0,\bf{k})\simeq \mathcal{N}
\begin{pmatrix}
\frac{k_i \sigma^i}{k^{1-2m_f}} &~& 0\\
\vspace{-.3cm}\\
0 &~&-\frac{k_i\sigma^i}{k^{1-2 m_f}}
 \end{pmatrix},
\ee where $\mathcal{N}=\frac{\Gamma[1/2-m_f]}{\Gamma[1/2+m_f]4^{m_f}}$ is an overall normalization constant and ${\bf k}=(k_x,k_y,k_z)$. Note that when $m_f$ is negative, the Green function has poles at $\omega=k$ while when $m_f$ is positive, the Green function has zeros instead of poles at $\omega=k$. However, from the procedure below, we will see that the topological structure %will not be 
is not 
affected by the value of the scaling dimension and no matter whether the Green function has zeros or poles, topological invariants could be the same.

The topological Hamiltonian $\mathcal{H}_t$ is defined as $-G^{-1}(0,\bf{k})$ from (\ref{eq:topH}). To calculate the Berry curvature, we need to find eigenvectors of the topological Hamiltonian which are equivalent to eigenvectors of the Green function. For the pure AdS case, the eigenvalues of the Hamiltonian are degenerate at $\omega=0$. Here we can treat this system as a $b\to 0$ limit of the Weyl semimetal case, where two Weyl points join to form a Dirac point. Then we could separate the two eigenvectors according to their chiralities, i.e. the two eigenvectors with negative eigenvalues in fact correspond to the two eigenvectors of the two chiral Hamiltonians with only one chirality of spinor each. The two eigenvectors are then
\be
| n_1\rangle=n_1^0\big(k_z+k,k_x+i k_y,0,0\big)^T\,,~~~~
| n_2\rangle=n_2^0\big(0,0,k_z-k,k_x+i k_y\big)^T\,,
\ee  
where $n_{l}^0=1/\sqrt{2k(k-(-1)^l k_z)}$. Note that the eigenvectors for the pure AdS case are in fact the same as those in the free massless Dirac Hamiltonian. $|n_1\rangle$ has positive chirality and is the eigenvector of the positive chirality Hamiltonian while $|n_2\rangle$ has negative chirality and is the eigenvector of the negative chirality Hamiltonian. 

To calculate the topological invariant we define a sphere ${\bf S}: k=k_0$ enclosing the Dirac node $k=0$ where $k_0$ is a constant. The system is gapped on the sphere and the formula for the topological invariant is
\be
C_l=\frac{1}{2\pi}\oint_S {\bf \Omega}_l \cdot d{\bf S}\,,
\ee where \be
\Omega^i=\epsilon^{ijk }\mathcal{F}_{ij}\,,~~~~~~{\text{with}}~(i\,,j\,,k)\in\{k_x\,,k_y\,,k_z\}
\ee and $\mathcal{F}$ is the Berry curvature defined in (\ref{eq:berrycur}). $C$ defined in this way is an integer number that does not depend on the exact shape and radius of ${\bf S}$ as long the deformation does not pass through a Dirac node. 

On the sphere ${\bf S}=k_0(\sin\theta\cos\phi,\,\sin\theta\sin\phi,\,\cos\theta)$ we have %\comment{check}
${\bf \Omega}_l=(-1)^{l}{\bf e_\rho}/2k_0^2$\,, %\,,~~~~{\bf S}=a_0(\cos\theta\sin\phi,~\sin\theta\sin\phi,~\cos\phi)\,,
thus for $| n_1\rangle$ 
\be
C_1=\frac{1}{2\pi}\oint_S {\bf \Omega} \cdot d{\bf S} =\frac{1}{2\pi}\int_0^{2\pi}d\phi \int_0^\pi  d\theta \sin\theta\,  k_0^2 \frac{-1}{2k_0^2 }=-1
\ee 
while for $| n_2\rangle$
\be
C_2=\frac{1}{2\pi}\oint_S {\bf \Omega} \cdot d{\bf S}=\frac{1}{2\pi}  \int_0^{2\pi}d\phi \int_0^\pi d\theta \sin\theta\,  k_0^2 \frac{1}{2k_0^2 }=1\,.
\ee The total topological invariant is then zero for pure AdS. This is clear intuitively: the dual zero density state of pure AdS$_5$ only consists massless Dirac excitations. %\comment{check which is -1 which is 1}

%%%%%%%%%%%%%%%%%%%%%%%%%%%%%%%%%%%%%%%%%%%
\subsection{Topological invariant for the holographic Weyl semimetal}
%%%%%%%%%%%%%%%%%%%%%%%%%%%%%%%%%%%%%%%%%%%

For the Weyl semimetal, the nontrivial topological invariant is defined as the Berry curvature integrated on a closed surface ${\bf S}$ enclosing the Weyl node located at ${\bf k}_l$ in the momentum space
\be
C^{\text{Weyl}}_l=\frac{1}{2\pi}\oint {\bf \Omega}_l\cdot d{\bf S}\,,
\ee  
and the result does not depend on the exact shape and size of ${\bf S}$ as long as there is only one Weyl node inside the closed surface. %\comment{A picture of the sphere to calculated the berry curvature, not so necessary}

For the Weyl semimetal case, the zero frequency Green function, or equivalently the effective topological Hamiltonian is also real. We will start from the easiest case: the $M/b \to 0$ limit where the contribution of $\phi$ is infinitely small so that could be ignored. Then we go to the more general case of small $M/b$. This $M/b \to 0$ limit is also a probe limit which is valid for the holographic Weyl semimetal away from the quantum critical point.
\subsubsection{$M/b\to 0$ limit}

In the $M/b\to 0$ limit, we ignore the backreaction of $\phi$ to the background geometry and to the axial gauge field. Then the axial gauge field is a constant in the bulk with $A_z=a_0$ %, which we denote as $a_0$ and the background is 
and the metric is pure $AdS_5$. As we ignore the contribution of $\phi$, $\psi_1$ and $\psi_2$ do not couple together and could be solved independently in terms of BesselK functions at $\omega =0$, 
%\be
%\psi_{1,2}^+=r^{-1/2}K_{\pm m+\frac{1}{2}}\Big(\frac{k_{1,2}}{r}\Big)a^+_{1,2}\,,
%\ee  
\be
\psi_{l}^+=r^{-1/2}K_{-(-1)^l m_f+\frac{1}{2}}\Big(\frac{k_{l}}{r}\Big)a^+_{l}
\ee
where ${\bf k}_{l}=(k_x,k_y,k_z+(-1)^la_0)$ and $k_l=\sqrt{k_x^2+k_y^2+(k_z+(-1)^la_0)^2}$. %while $k_{2i}=(k_x,k_y,k_z+a_0)$ and $k_2=\sqrt{k_x^2+k_y^2+(k_z+a_0)^2}$.
Compared to the pure AdS case, the pole of the first spinor (the negative chirality one) shifts from $w={\bf k}=0$ to $w=k_x=k_y=0$ while $k_z=a_0$ and the pole of the second spinor shifts to $k_z=-a_0$.  The retarded Green function is
\be\label{weylgreenM0}
G(0,\bf{k})\simeq \mathcal{N}
\begin{pmatrix}
\frac{k_{1i} \sigma^i}{k_1^{1-2m_f}} &~& 0\\
\vspace{-.3cm}\\
0 &~&-\frac{k_{2i}\sigma^i}{k_2^{1-2 m_f}}
\end{pmatrix},
\ee
where the normalisation factor $\mathcal{N}$ takes the same form as in (\ref{eq:specads}). 
The eigenvalues of the Green function give $\pm \mathcal{N} k_{1,2}^{2m_f}$. For negative $m_f$ the poles of the system are at $k_{1,2}=0$. As $a_0$ is not zero, the eigenvectors are now not degenerate at zero frequency. At the two Weyl nodes $k_z=\pm a_0$, i.e. one of $k_{1,2}$ is zero while the other not zero, two branches of the eigenvectors are gapless and the other two are gapped. 

Now we calculate the topological invariant at $k_z=a_0$ and the calculation for the other one would be similar and give an opposite topological invariant.
At  $k_z=a_0$, the gapless eigenvector with a negative eigenvalue of the topological Hamiltonian is 
\be
|n_1\rangle =n_0\big(k_z-a_0+k_1,k_x+i k_y,0,0\big)^T\,,
\ee  
where $n_0=1/\sqrt{2k_1(k_1+k_z-a_0)}$ with $k_1=\sqrt{k_x^2+k_y^2+(k_z-a_0)^2}$. 
From $|n_1\rangle$ we have ${\bf{\Omega}}=-{\bf{e}}_\rho/2k_1^2$ and the topological invariant is 
\be
C^{\text{Weyl}}_1=\frac{1}{2\pi}\oint_{\bf S} {\bf \Omega}_1\cdot d{\bf S} = 
\frac{1}{2\pi}\int_0^{2\pi}d\phi \int_0^\pi d\theta  \sin\theta\, k_1^2 \frac{-1}{2k_1^2 }=-1\,.
\ee
It can be checked that the gapped eigenvector will only contribute a zero to the topological invariant.

This shows that $k_z=a_0$ is a Weyl node with negative 
chirality and the other node at $k_z=-a_0$ should have $C^{\text{Weyl}}_2=1$, i.e. the other node possesses an opposite chirality and topological charge because the total topological invariant/chirality charge for the whole systems should still be zero which is exactly the consequence of the Nielsen-Ninomiya theorem \cite{Nielsen}. 

This is the simplest case that $\phi$ does not have any contribution. The next step is to calculate the topological invariant for the more general nonzero $M/b$ case. For this case, the background geometry gets modified by the scalar field in the bulk and we cannot solve it analytically anymore. As the background is numerical, we do not have analytic solutions for the fermion Green function either. We could in principle calculate the retarded Green functions using numerics, and then calculate the eigenstates and Berry curvature using numerics. However, this procedure requires finding eigenvectors numerically which usually loses a lot of accuracy. In order to avoid too much numerics and the inaccuracy, we will solve it semi-analytically by expanding near the Weyl nodes and for this to be possible we have to work at the small $M/b\ll (M/b)_c$ limit.

%%%%%%%%%%%%%%%%%%%%%%%%%%%%%%%%%%%%%%%%%%%
\subsubsection{$M/b\ll (M/b)_c$ case}
\label{secinapp}
%%%%%%%%%%%%%%%%%%%%%%%%%%%%%%%%%%%%%%%%%%%

For $M/b\ll (M/b)_c$, it is expected that there would be two poles separated in the $k_z$ axis, though close to each other. 
When we calculate the Berry curvature we could in principle perform the integration on any sphere that surrounds one and only one Weyl node.  When we reduce the size of the sphere to be smaller and smaller, we could expand the system around the Weyl node to solve the fermions and get the retarded Green function on that sphere and then diagonalize it for the eigenstates. Here because of the non-analyticity in the equations of the probe fermions near the pole, we need to take a near-far matching method.

\begin{figure}[h!]
\begin{center}
\includegraphics[width=0.63\textwidth]{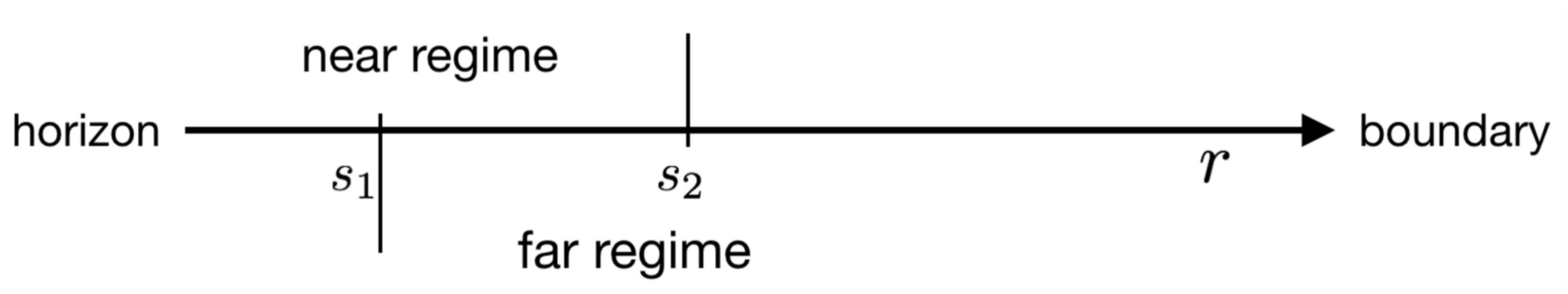}
\end{center}
\vspace{-0.3cm}
\caption{\small Illustration of the near region $r\ll s_2$ and the far region $s_1\ll r$.}
\label{fig:nf}
\end{figure}

We can divide the geometry into the near region and the far region, which overlap at the matching region. There are usually two scales $s_1\ll s_2$ where the near region is defined by $r\ll s_2$ and far region $s_1\ll r$ while the matching region $s_1\ll r\ll s_2$ as illustrated in Fig. \ref{fig:nf}. $s_1$ is the IR expansion parameter which is  important in the near region while not important in the far region. Here in this system $s_1$ is $k_1$ if we focus on the right Weyl node at $k_z=a_0$ and the $k_x$, $k_y$ and $k_z-a_0$ terms in the far region could be treated as perturbations. $s_2$ is a UV parameter, e.g. $s_2$ is the chemical potential $\mu$ in the finite density case. Here $s_2$ is the parameter at which the geometry starts to deviate from $AdS_5$. According to the background geometry (\ref{eq:wsm-nh}) $s_2$ is in fact $a_1$ or equivalently $b$ as $a_1/b \sim O(1)$. The near region is now $r\ll b$ and in this region the background is $AdS_5$. Thus we require $k_1=\sqrt{k_x^2+k_y^2+(k_z- a_0)^2}\ll b$ for the near-far matching method to work. %in the Weyl system.

In the near region of the holographic Weyl semimetal phase, the contribution of $\phi$ almost vanishes while the leading order of $A_z$ is a constant. The near horizon geometry is still $AdS_5$. As the order of $\phi$ is extremely small at the horizon in the Weyl semimetal phase, the equations for the two spinors are decoupled in the near region. We calculate near one of the expected poles and choose $k_{z0}=a_0$.

The near region equations are
\be
\Bigg(\Gamma^{\underline{r}}\partial_r+\frac{1}{u}\Big(-i\omega \Gamma^{\underline{t}}+ik_x\Gamma^{\underline{x}}+ik_y\Gamma^{\underline{y}}\Big)+\frac{1}{\sqrt{uf}}\Big(i(k_z+(-1)^l A_z) \Gamma^{\underline{z}}\Big)
+(-1)^l \frac{m_f}{\sqrt{u}}\Bigg)\psi_{l}=0\,,\nn
\ee
and at the near region the solutions are 
\be\label{eq:nh1}
\psi_{l}^+=r^{-1/2}K_{-(-1)^l m_f+\frac{1}{2}}\Big(\frac{k_{l}}{r}\Big)a^+_{l}\,,
\ee  
where $k_l=\sqrt{k_x^2+k_y^2+(k_z+(-1)^l a_0)^2}$ 
for the upper two components, and using 
\be
\psi^-_{l}=\frac{k_{l\mu}\sigma^\mu}{k_{l}^2}r\big(r\partial_r+(-1)^l m_f\big)\psi^+_{l}
\ee 
we get
\be\label{eq:nh2}
\psi^-_{l}=\frac{k_{l\mu}\sigma^\mu}{k_{l}} r^{-1/2} K_{-(-1)^l m_f-\frac{1}{2}}\Big(\frac{k_{l}}{r}\Big)a^+_{l}
\ee 
for the lower two components.

The far region equations can be expanded in terms of ${\bf k}_1=(k_x, k_y, k_z-a_0)$ around ${\bf 0}$ as
\be
\psi_{l}^f=\psi_{l}^{f0}+k_{1 i} \psi_{l}^{f1 i}\,,
\ee 
where $k_x,~k_y,~k_z-a_0$ are the small expansion parameters. The far region leading order equations are
\be\label{eq:farlead}
\Bigg(\Gamma^{\underline{r}}\partial_r+\frac{1}{\sqrt{uf}}\Big(i(a_0\mp A_z) \Gamma^{\underline{z}}\Big)
+(-1)^l\frac{m_f}{\sqrt{u}}\Bigg)\psi_{l}^{f0}-\eta_1\frac{\phi}{\sqrt{u}}\psi_{\bar{l}}^{f0}=0\,,
\ee 
and to the first order in $k_{1i}$ equations are
\be
\Bigg(\Gamma^{\underline{r}}\partial_r+\frac{1}{\sqrt{uf}}\Big(i(a_0\mp A_z) \Gamma^{\underline{z}}\Big)
+(-1)^l\frac{m_f}{\sqrt{u}}\Bigg)\psi_{l}^{f1j}-\eta_1\frac{\phi}{\sqrt{u}}\psi_{\bar{l}}^{f1j}+\frac{1}{u}\Big(ik_{j}\Gamma^{\underline{j}}\Big)\psi_{\bar{l}}^{f0}=0\nn
\ee 
for $k_j\in \{ k_x, k_y\}$, and
\be
\Bigg(\Gamma^{\underline{r}}\partial_r+\frac{1}{\sqrt{uf}}\Big(i(a_0\mp A_z) \Gamma^{\underline{z}}\Big)
+(-1)^l \frac{m_f}{\sqrt{u}}\Bigg)\psi_{l}^{f1z}-\eta_1\frac{\phi}{\sqrt{u}}\psi_{\bar{l}}^{f1z}+\frac{1}{\sqrt{uf}}\Big(i(k_z-a_0) \Gamma^{\underline{z}}\Big)\psi_{l}^{f0}=0\nn
\ee 
for $k_z-a_0$, where $\bar{l}=3-l$.

To solve the far region equations we need the near horizon boundary conditions which are input determined by the matching region expansion of the near region. In the near horizon region of the far region, i.e. at the matching region, as $\phi$ is %still 
not important from (\ref{eq:farlead}) the leading order solutions are %\comment{CHECK}
\be
\psi_{1}^{f0}=\begin{pmatrix}
%r^{\pm m_f}
r^{ m_f} a_{1}^+\\
\vspace{-.3cm}\\
%r^{\mp m_f}
r^{- m_f} a_{1}^-
\end{pmatrix}
\ee for $\psi_{1}$ and
\be
\psi_{2}^{f0}=\begin{pmatrix}
r^{-1/2}K_{- m_f+\frac{1}{2}}\Big(\frac{2 a_0}{r}\Big)a^+_{2}\\
\vspace{-.3cm}\\
r^{-1/2} K_{- m_f-\frac{1}{2}}\Big(\frac{2 a_0}{r}\Big)\sigma^z a^+_{2}
\end{pmatrix}
\ee for $\psi_{2}$, where $a_1^{\pm}$ and $a_{2}^{+}$ are constant two-component spinors.  There are six independent near horizon parameters $a_1^{\pm}$ and $a_{2}^{+}$, and to calculate the retarded Green function, we only need four nontrivial linearly independent combinations of the six, which are determined by the boundary conditions at the matching region. Note that for $\psi_2$ there are only two free parameters compared to four for $\psi_1$. This is because the expansion around $k_z=a_0$ is analytic for $\psi_2$ and the infalling boundary conditions for $\psi_2$ have already been chosen in the far region. 

At first order in $k_{1i}$ of the $i$-th component in ${\bf k}_1=(k_x, k_y, k_z-a_0)$, the solutions are sourced by the leading order solutions and there are no new free parameters. We subtract all the solutions of the leading order and the near horizon solutions are only nonzero when there are nonzero leading order sources. The first order solutions are %\comment{CHECK}
\be
\psi_{1}^{f1i}=\begin{pmatrix} 
r^{-1- m_f} c_{1i}^+\\
\vspace{-.3cm}\\
r^{-1+ m_f}c_{1i}^-\end{pmatrix}\,,~~~\text{with~} i\in\{x,y,z\}
\ee where $c_{1x,y,z}^{+,-}$ are two component spinors that are determined by $a_1^{\pm}$ and $a_{2}^{+}$.

\be
\psi_2^{f1i}=r^{-\frac{3}{2}}\begin{pmatrix} 
K_{m_f+\frac{3}{2}}\left(\frac{2 a_0}{r}\right)c_{2i}^+\\
\vspace{-.3cm}\\
K_{-m_f+\frac{1}{2}}\left(\frac{2 a_0}{r}\right)c_{2i}^-\end{pmatrix}\,,~~~\text{with~} i\in\{x,y,z\}
\ee  
where $c_{2x,y,z}^{+,-}$ are two component spinors that are determined by $a_1^{\pm}$ and $a_{2}^{+}$.

Before going to the matching region to match the initial conditions of $\psi_1$, which should be two sets of linearly independent combinations of $\begin{pmatrix} a_{1}^{+} \\ a_{1}^{-} \end{pmatrix}$
, we could first obtain the boundary values of the fields under these six independent far region boundary conditions. For simplicity we choose these six boundary conditions to be 
\be
V_{\text{ini},i}^j=\delta_i^j\,,~~~\text{with~} i,\,j\in\{1,...6\}\,,
\ee where $V_{\text{ini}}=\begin{pmatrix} 
a_{1}^{+}\\
a_{1}^{-}\\
%\vspace{-.3cm}\\
a_{2}^{+}
\end{pmatrix}
%\big( a_{1}^{+}\,,~a_{1}^{-}\,,~a_{2}^{+}\big)^T
$ and $V_{\text{ini},i}^j$ refers to the value of the $i$-th component of $V_{\text{ini}}$ under the $j$-th boundary condition.

Now we indicate the boundary source vector as $s_{i}^{j}$ and the expectation vector as $e_{i}^j$, which are the $i$-th components of $r^{-m_f}\begin{pmatrix}\psi_1^+\\ \psi_2^-\end{pmatrix}$ and $r^{m_f}\begin{pmatrix}-\psi_2^+\\ \psi_1^-\end{pmatrix}$ under the $j$-th boundary condition separately. We keep terms in both matrices $s$ and $e$ up to the first order in $k_x, k_y$ and $k_{z}-a_0$ and each element of $s_{i}^j$ or $e_i^j$ would be composed of zeroth and first order contributions in $k_x, k_y$ and $k_{z}-a_0$. Due to the structure of the equations, some of these contributions would be zero, e.g. $s_1^1$ has no $O(k_x)$ or $O(k_y)$ contributions. A full set of nonzero elements of these two matrices $s$ and $e$ under the six boundary conditions could be found in the appendix \ref{appA}. The exact values of these elements could be obtained by numerically integrating the far region equations for any background in the holographic Weyl semimetal phase.

In the matching region, $k_1\ll r\ll b$ with $k_1=\sqrt{k_x^2+k_y^2+(k_z-a_0)^2}$ we expand the near region solutions (\ref{eq:nh1}) and (\ref{eq:nh2}) as
\bea
\psi_1^+&=&\left[r^{m_f}\left(2^{m_f-\frac{1}{2}} k_1^{-m_f-\frac{1}{2}} \Gamma \left(m_f+\frac{1}{2}\right)\right)+r^{-m_f-1} \left({2^{-m_f-\frac{3}{2}} k_1^{m_f+\frac{1}{2}} \Gamma \left(-m_f-\frac{1}{2}\right)}\right)\right] d_1^+\,\nn\\
&&\nn\\
\psi_2^+&=&\left[\frac{1}{\sqrt{r}}K_{m_f+\frac{1}{2}}\left(\frac{2 h_0}{r}\right)+\frac{(k_2-2 a_0) }{4 r^{3/2}}\left(\frac{(2 m_f +1)r }{a_0}K_{m_f+\frac{1}{2}}\left(\frac{2 a_0}{r}\right)-4 K_{m_f+\frac{3}{2}}\left(\frac{2 a_0}{r}\right)\right) \right]d_2^+\nn
\eea
for the upper two components, and
\bea
\psi_{1}^-&=&\left[2^{m_f-\frac{3}{2}} k_1^{-m_f-\frac{1}{2}} r^{m_f-1} \Gamma \left(m_f-\frac{1}{2}\right)+2^{-m_f-\frac{1}{2}} k_1^{m_f-\frac{3}{2}} r^{-m_f} \Gamma \left(\frac{1}{2}-m_f\right)k_{1\mu}\sigma^{\mu}\right] d_1^+\nn\\
&&\nn\\
\psi_{2}^-&=&\bigg[\frac{1}{2 a_0\sqrt{r}}K_{-m_f-\frac{1}{2}}\left(\frac{2 a_0}{r}\right)+\frac{(k_2-2 a_0)}{8 a_0^2 r^{3/2}}\Big(-4a_0 K_{-m_f+\frac{1}{2}}\left(\frac{2 a_0}{r}\right)\nn\\
&&\nn\\
&&~~~~-(2m_f+3) r K_{-m_f-\frac{1}{2}}\left(\frac{2 a_0}{r}\right)\Big)\bigg] d_2^+\nn
\eea for the lower two components.

These expansions fix the near horizon initial boundary conditions for the far region and we could solve the far region equations using these boundary conditions and obtain the source and expectation matrices under infalling boundary conditions. In fact for $\psi_2$ we do not need this near far matching procedure and could directly use the infalling boundary conditions at the far region and treat $k_1$ as a small expansion as it is analytical when expanding around $k_{2}^0=2 a_0$ which is not zero.

Now the four infalling boundary conditions that we need are %\comment{CHECK}
\be\label{match}
\begin{pmatrix} 
d_{1}^{+j}\\
\vspace{-.3cm}\\
d_{2}^{+j}
\end{pmatrix}_i=\delta_i^j
\ee
%\be\label{match}
%\left(d_1^+, d_2^+ \right)^T\in \big{\{}(1,0,0,0)^T\,,~(0,1,0,0)^T\,,~(0,0,1,0)^T\,,~(0,0,0,1)^T\big{\}}\,,
%\ee 
%or equivalently ${\left(d_1^+, d_2^+ \right)^T}_i^j=\delta_i^j$, 
i.e. 
the $i$-th component of $\begin{pmatrix} 
d_{1}^{+}\\
d_{2}^{+}
\end{pmatrix}$ under the $j$-th boundary condition is $\delta_i^j$. These four boundary conditions fix the six far region boundary conditions to be four and the boundary values of the fields under these four boundary conditions are also combinations of the boundary values of the fields under six far region boundary conditions. Let us denote $s^j$ and $e^j$ as the source and expectation vectors under the far region $j$-th boundary condition, and $j$ runs from $1,...,6$. The source vector for the first matching region boundary condition in (\ref{match}) corresponds to
\be
\left( \Gamma \left(m_f+\frac{1}{2}\right)\frac{2^{m_f-\frac{1}{2}}}{ k_1^{m_f+\frac{1}{2}}}\right) (s^1)+(s^3,s^4)\cdot \left(\Gamma \left(\frac{1}{2}-m_f \right)\frac{ k_1^{m_f-\frac{3}{2}}}{2^{m_f+\frac{1}{2}}}k_{1\mu}\sigma^{\mu}\begin{pmatrix}1\\ 0\end{pmatrix}\right),
\ee 
and the source vector for the second matching region boundary condition in (\ref{match}) corresponds to
\be
\left(\Gamma \left(m_f+\frac{1}{2}\right)\frac{2^{m_f-\frac{1}{2}}}{ k_1^{m_f+\frac{1}{2}}}\right) (s^2)+(s^3,s^4)\cdot \left(\Gamma \left(\frac{1}{2}-m_f \right)\frac{ k_1^{m_f-\frac{3}{2}}}{2^{m_f+\frac{1}{2}}}k_{1\mu}\sigma^{\mu}\begin{pmatrix}0\\ 1\end{pmatrix}\right).
\ee The third and fourth boundary conditions correspond to $s^5$ and $s^6$ separately. For expectations, we only need to substitute $s$'s in the formulas above by $e$'s.

Finally we could get the source and expectation matrices at the boundary, which are composed of the parameters $x_{i}$ in the appendix \ref{appA}, where $x_i$ with $i\in 1, ..., 36$ are constants which are the boundary values of the solutions associated with the six far region boundary conditions. These two matrices are very long and we do not write them out here. The next step is to get the Green function from the source and expectations matrices using $G=i\Gamma_t e s^{-1}$. The Green function obtained in this way is still quite complicated and it is difficult to obtain the eigenstates of the Green function. Now we analyze the Green function more carefully to see if it could be simplified in certain limits.

 As the final Green function will not get modified by changing the initial boundary conditions by a linear superposition or scaling, here we rescale both the source and expectation matrix by a factor of $\sqrt{k_1}$ for simplicity. Now the determinant of the source matrix is 
\bea\label{det1}
\det S&=&\frac{2  (x_{16}x_{25}-x_{13}x_{28})^2}{\Gamma \left(\frac{1}{4}\right)^2/\Gamma \left(\frac{3}{4}\right)^2}+\frac{2 \sqrt{2} k_{1z}(x_{1} x_{28}-x_{25} x_{4}) (x_{16} x_{25}-x_{13} x_{28})}{\sqrt{k_1} \Gamma \left(\frac{1}{4}\right)/ \Gamma \left(\frac{3}{4}\right) }+
\nn\\
&&\nn\\
&&~~+t_1(x_{i})k_{1z}+t_2(x_{i}) k_1\,,
\eea  
written in orders of $k_1^0,k_1^{1/2},k_1$, where $t_1$ and $t_2$ are functions of $x_{i}$ which are too long to write out. As we have stated above, $x_{i}$ with $i\in 1,...,36$ are 36 constants that could be read from the boundary values of the source and expectation matrices and $k_1=\sqrt{k_x^2+k_y^2+(k_z-a_0)^2}$. Note that $t_{1,2}$ do not have the factor $(x_{16}x_{25}-x_{13}x_{28})$.

In the pure AdS case and the $M/b\to 0$ limit, as $\psi_1$ and $\psi_2$ do not couple together it can be checked that $x_{16} x_{25} - x_{13} x_{28}=0$. In this limit $\det S$ could be simplified to
\be\label{det2}
\det S\simeq (x_1 x_{28}-x_{25} x_{4})^2 k_1\,.
\ee The Green function is also %would also be %much simpler 
simplified in this limit. For $M/b$ small enough, %we would show that 
in principle one could find an enclosing sphere with radius $k_1$ on which $x_{16} x_{25} - x_{13} x_{28}\ll (x_1 x_{28}-x_{25} x_{4}) \sqrt{k_1}$ holds and the expressions for the Green function and the topological invariants could be simplified a lot as small perturbations would not change the topological invariants. $x_i$'s are parameters that do not depend on $k_1$ so it seems that if we choose $k_1$ large enough, this inequality would hold. However, in our near far matching calculation, the order of $k_1$ at the sphere should be so small that there exists a region of $r\gg k_1$ where the system could still be AdS$_5$ and also that $2 k_1$ at the enclosing sphere should be smaller than $2 a_0$ so that the sphere only has one pole inside. With the largest possible $k_1$ that satisfies this constraint, we find that $x_{16} x_{25} - x_{13} x_{28}\ll (x_1 x_{28}-x_{25} x_{4}) \sqrt{k_1}$ indeed holds for small values of $M/b$ in the holographic semimetal phase. Numerically we have checked that for $M/b\simeq 0.16$, the ratio of the left side of the inequality over the right side could be around $7.7\%$ at the sphere where we have chosen $k_1=10^{-2} a_0$ or if we choose $k_1=10^{-4} a_0$ the ratio would be around $6.7\%$ for $M/b\simeq 0.05$. Here we have chosen $m_f=-1/4$ without loss of generality.

Note that the position of poles should be at $\det S=0$ and from the perturbative calculation (\ref{det1}) the position of the pole seems also to be modified for a very small value compared to $a_0$ in the case of $M/b\to 0$ (\ref{det2}) due to the $x_{16} x_{25} - x_{13} x_{28}$ term, which though could be ignored in the limit that we are considering. However, with a nonzero $\omega\to 0$ we find that this pole is  not visible in the imaginary part of the Green function as the Green function is purely real at $\omega\to 0$ while $k_z\neq a_0$. This means that for the imaginary part, the peak is still at $a_0$. Note that the value of anomalous Hall conductivity $\sigma_\text{AHE}\simeq 8\alpha a_0$ should be proportional to the distance between two poles, which seems to also lead to the conclusion that the position of poles should be at $\pm a_0$. Thus it is possible that there is some reason leading to the fact that summing over all perturbations at the order $x_{16} x_{25} - x_{13} x_{28}$ would finally keep the position of the pole unchanged. We will leave this for a future study.

We have found that at the sphere with a small radius $k_1$, which is away from the pole but not far away, $ (x_{16}x_{25}-x_{13}x_{28})$ is very small compared to other terms as we focus on the case that $M/b$ is small enough. Thus the $ (x_{16}x_{25}-x_{13}x_{28})$ term could be ignored as the Berry curvature is a quantized number which should not be affected by small perturbations. In this limit, %we could get 
 the Green function can be simplified to be 
%\be
% G(0,{\bf k})\simeq \frac{N_1}{ \det S}\left(
%\begin{array}{cccc}
% \frac{x_{28} k_{1z}}{x_{25} \sqrt{k_1}} & \frac{x_{28} (k_x-i k_y)}{x_{25} \sqrt{k_1}} & -\frac{ k_{1z}}{\sqrt{k_1}} & \frac{k_x-i k_y}{\sqrt{k_1}} \\
% \vspace{-.3cm}\\
% \frac{x_{28} (k_x+i k_y)}{x_{25} \sqrt{k_1}} & -\frac{x_{28}  k_{1z}}{x_{25} \sqrt{k_1}} & \frac{-k_x-i k_y}{\sqrt{k_1}} & -\frac{ k_{1z}}{\sqrt{k_1}} \\
% \vspace{-.3cm}\\
%N_2 \frac{x_{28} k_{1z}}{x_{25} \sqrt{k_1}} & N_2 \frac{x_{28} (k_x-i k_y)}{x_{25} \sqrt{k_1}} &-N_2 \frac{ k_{1z}}{\sqrt{k_1}} & N_2\frac{ k_x-i k_y}{ \sqrt{k_1}} \\
%\vspace{-.3cm}\\
% -N_2\frac{x_{28} (k_x+i k_y)}{x_{25} \sqrt{k_1}} &N_2 \frac{x_{28} k_{1z}}{x_{25} \sqrt{k_1}} & N_2\frac{ (k_x+i k_y)}{ \sqrt{k_1}} &N_2 \frac{ k_{1z}}{ \sqrt{k_1}} \\
%\end{array}
%\right),
%\ee 
\be
 G(0,{\bf k})\simeq \frac{N_1}{\sqrt{k_1} \det S}\left(
\begin{array}{cccc}
N_0 k_{1z} & N_0(k_x-i k_y) & - k_{1z} & k_x-i k_y \\
 \vspace{-.3cm}\\
N_0 (k_x+i k_y)& -N_0  k_{1z} & -k_x-i k_y & - k_{1z} \\
 \vspace{-.3cm}\\
N_0 N_2  k_{1z}& N_0 N_2  (k_x-i k_y)&-N_2 k_{1z} & N_2( k_x-i k_y) \\
\vspace{-.3cm}\\
 - N_0 N_2 (k_x+i k_y) &N_0 N_2 k_{1z} &N_2 (k_x+i k_y)&N_2  k_{1z} \\
\end{array}
\right)\,,\nn
\ee 
where $$N_0=\frac{x_{28}}{x_{25}}\,,~  
N_1=\sqrt{2}\big(x_{1} x_{28}-x_{25} x_{4}\big) \big(x_{22}x_{25}-x_{13} x_{34}\big)\frac{\Gamma \left(\frac{3}{4}\right)}{\Gamma \left(\frac{1}{4}\right)}~\text{and}~N_2=\frac{ x_{19} x_{25}-x_{13} x_{31} }{ x_{22} x_{25}-x_{13} x_{34} } .$$

From numerics one can check $N_2<N_0$ for $M/b\ll 1.$ 
We pick the negative valued normalized eigenstate of the topological Hamiltonian, which is %\comment{CHECK}
\be
\frac{1}{N}\left(\frac{-k_1- k_{1z} }{(k_x+i k_y) N_2}\,,~~-\frac{1}{N_2}\,,~~\frac{-k_1- k_{1z} }{k_x+i k_y}\,,~~1\right)^T
\ee 
%or
%\be
%\frac{1}{N}\left(\frac{k_1- k_{1z} }{(k_x+i k_y) N_2}\,,~~-\frac{1}{N_2}\,,~~\frac{k_1- k_{1z} }{k_x+i k_y}\,,~~1\right)~~~
%\ee 
with $N=\frac{\sqrt{2k_1(1+N_2^2)(k_1+k_z)}}{N_2\sqrt{k_x^2+k_y^2}}$. With this state, we could calculate the Berry curvature, then integrate it on the small enclosing sphere and get a nontrivial topological invariant $-1$. We could do a similar analysis for the pole at $k_z=-a_0$ and obtain the topological invariant $1$.  

Thus we have the final result for the nontrivial topological invariants for small $M/b$ for the holographic Weyl semimetal phase. For $M/b\sim O((M/b)_c)$ this matching method could not work as there is no matching region anymore. In this case we could in principle calculate the zero frequency Green functions using the numerical method and perform the integration also numerically to get topological invariants, which we leave for future investigation. 

%%%%%%%%%%%%%%%%%%%%%%%%%%%%%%%%%%%%%%%%%%%
\subsubsection{Spectral function at $\omega =\pm a_0$ and $k_z=0$}
%%%%%%%%%%%%%%%%%%%%%%%%%%%%%%%%%%%%%%%%%%%

We have calculated the topological invariant for the holographic Weyl semimetal using the $\omega=0$ Green function (topological Hamiltonian) in the previous subsection. In this subsection, for completeness we will have a look at the Fermi spectrum for $\omega\neq 0$ and to avoid tedious numerics we will also stay in the semi-analytic regime of calculation. In the $M/b\to 0$ limit, when we calculate the Green functions at $\omega >k_{1,2}$, instead of the formula in (\ref{weylgreenM0}), we have 
\be
G(\omega,{\bf k})\simeq
\begin{pmatrix}
\frac{\omega+k_{1\mu} \sigma^\mu}{k_{1\omega}^{1-2m_f}} &~& 0 \\
\vspace{-.3cm}\\
0 &~&\frac{\omega-k_{2\mu}\sigma^\mu}{k_{2\omega}^{1-2 m_f}}
 \end{pmatrix},
\ee where $k_{l\omega}=\sqrt{\omega^2-k_x^2-k_y^2-(k_z+(-1)^l a_0)^2}$. The poles are at $k_{1\omega}=0$ or $k_{2\omega}=0$. At $k_z=0$ we could see that the two branches of $k_{1\omega}=0$ and $k_{2\omega}=0$ intersect at $k_z=0$ while $\omega=\pm \sqrt{k_x^2+k_y^2+a_0^2}$. This means that besides the two ``band crossing" points at $\omega=0$ and $k_z=\pm a_0$, we have another two ``band crossing" points at $\omega=\pm a_0$ and $k_z=0$ at $M/b\to 0$. The following figure shows the Fermi spectrum of the $M/b \to 0$ limit holographic Weyl semimetal. 
\begin{figure}[h!]
\begin{center}
\includegraphics[width=0.62\textwidth]{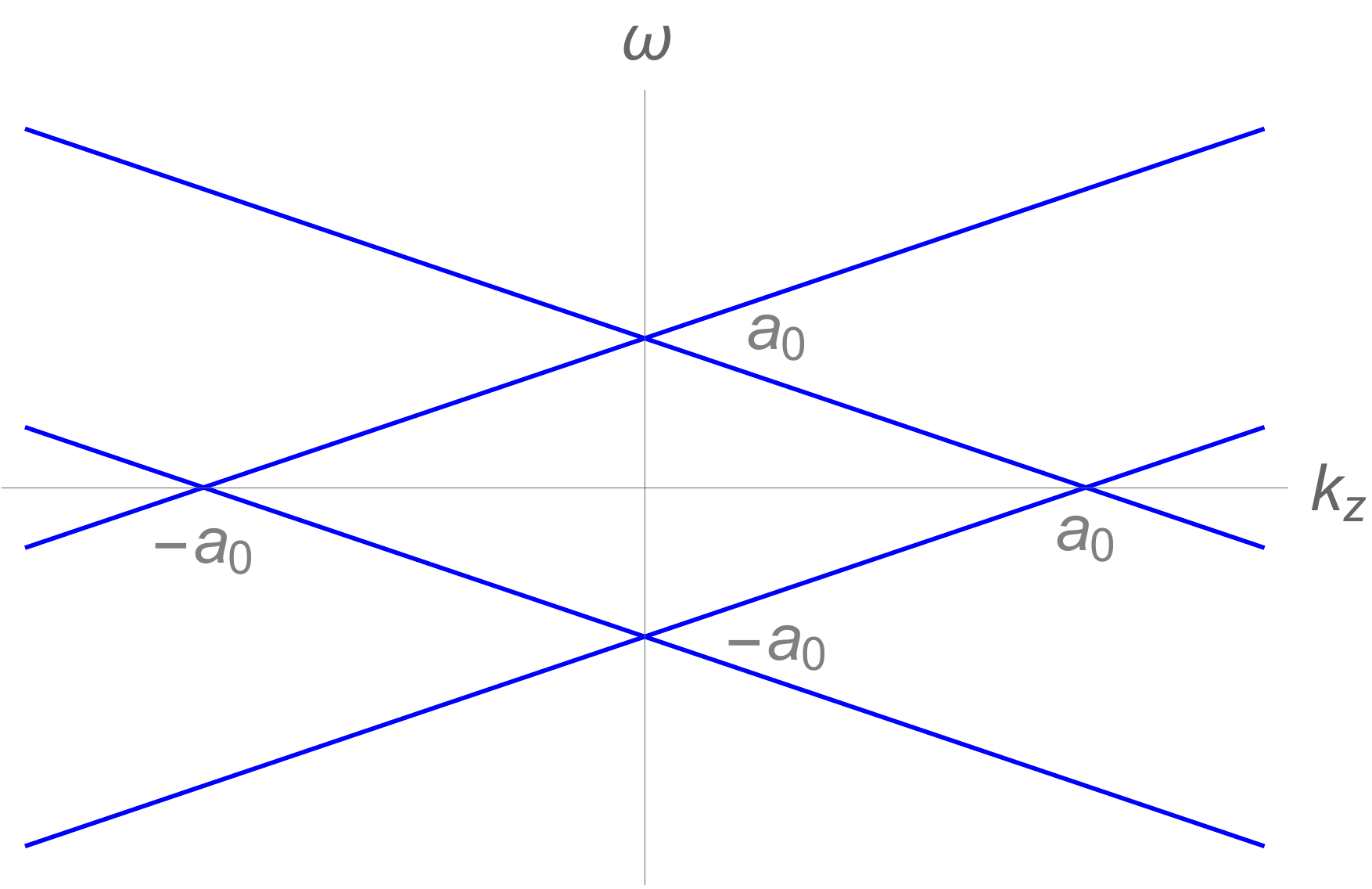}
\end{center}
\vspace{-0.4cm}
\caption{\small The spectrum of the holographic Weyl semimetal in the limit $M/b=0$. When we have a small nonzero $M/b$, the points $(0, \pm a_0)$ will become a pseudogap and we do not have poles at these two points any more. The points $(\pm a_0, 0)$ remain poles.}
\label{fig:npoints}
\end{figure}

In the following we will show that when $M/b\neq 0$, the effect of $\phi$ will change this band intersection at $\omega=\pm a_0$ and $k_z=0$ into a pseudogap. We work in the very small $M/b$ limit and expand the system in orders of $M/b$ to study the leading order effect of $M/b$. We could easily check from the equations of motion for the background that the scalar field has an order $O(M/b)$ profile and backreacts to other fields to give order $O\left((M/b)^2\right)$ order corrections to other fields. This means that at leading order in $M/b$ the background geometry would still be pure $AdS_5$ with $A_z=b=a_0$ all through the bulk and $\phi=\phi(r)\sim O(M/b)$ which could be solved from the equation of motion for $\phi$ in the AdS background. 

As the effect of $\phi$ at the horizon is always negligible in the holographic Weyl semimetal phase, again we take the near far matching method %and this time 
with 
$\omega-a_0\to 0$ and for simplicity also $k_x=k_y=0$. The near region is defined by $r\ll b$ and the far region is defined by $\omega\mp a_0\ll r$ depending on if we want to study the up or down branch. Here we focus on the $\omega\to a_0$ branch and it is straightforward to generalize to the other branch. 

In the near region, the geometry is $AdS$ and the near region solutions are the Hankel functions
\be
\psi_{1}=\begin{pmatrix} 
\frac{1}{\sqrt{r}}H^{(1)}\left(\frac{1}{4},\frac{k_{1\omega}}{r}\right)a^+_1\\
\vspace{-.3cm}\\
\frac{\omega+k_{1\mu} \sigma^{\mu}}{k_{1\omega}\sqrt{r}}H^{(1)}\left(-\frac{3}{4},\frac{k_{1\omega}}{r}\right)a^+_1
\end{pmatrix}\,,~~~~
\psi_{2}=\begin{pmatrix} 
\frac{1}{\sqrt{r}}H^{(1)}\left(\frac{3}{4},\frac{k_{2\omega}}{r}\right)a^+_2\\
\vspace{-.3cm}\\
\frac{\omega+k_{2\mu} \sigma^{\mu}}{k_{2\omega}\sqrt{r}}H^{(1)}\left(-\frac{1}{4},\frac{k_{2\omega}}{r}\right)a^+_2
\end{pmatrix}\,,~~~~
\ee
%\bea
%\psi_{1}^+&=&\frac{1}{\sqrt{r}}H^{(1)}\left(\frac{1}{4},\frac{k_{1\omega}}{r}\right)a^+_1\,,\\
%\psi_{1}^-&=&\frac{\omega+k_{1\mu} \sigma^{\mu}}{k_{1\omega}\sqrt{r}}H^{(1)}\left(-\frac{3}{4},\frac{k_{1\omega}}{r}\right)a^+_1\,,\\
%\psi_{2}^+&=&\frac{1}{\sqrt{r}}H^{(1)}\left(\frac{3}{4},\frac{k_{2\omega}}{r}\right)a^+_2\,,\\
%\psi_{2}^-&=&\frac{\omega+k_{2\mu} \sigma^{\mu}}{k_{2\omega}\sqrt{r}}H^{(1)}\left(-\frac{1}{4},\frac{k_{2\omega}}{r}\right)a^+_2\,,
%\eea
where $k_{l \omega}=\sqrt{\omega^2-k_x^2-k_y^2-(k_z+(-1)^l a_0)^2}$ and we have chosen $m_f=-1/4$.
Note that different from the $\omega=0,~k_z\to a_0$ region where $\psi_1$ has effectively zero momentum while $\psi_2$ has an effective finite momentum, here for each of $\psi_{l}$ half of the components have effectively zero momentum while the other half nonzero. However, the solutions are still functions of $k_{l\omega}/r$ where $k_{l \omega}\to 0$. This indicates that though for some components of $\psi_{l}$, there is a finite $\omega_0+a_0=2 a_0$ term in the equation, the $\omega_0+a_0=2 a_0$ terms in the equations are in fact first order $\omega-a_0$ corrections sourced by corresponding leading order solutions which have the $\omega+a_0$ coefficient in front. We define $\tilde{\omega}=\omega-a_0$ and fix $k_x=k_y=k_z=0$. We expand the system at $\omega\to a_0$ i.e. small $\tilde{\omega}$. As we explained above, to study the leading order effect of $M/b$ the background geometry is still $AdS_5$ and $A_z$ is also a constant in the whole bulk spacetime. Then we could also expand the equations of motion for $\psi_{l}$ in orders of $M/b$ at the far region as it is only important in the far region.

The far region solutions could be written as
\be
\psi_{l}^f=\psi_{l}^{f0}+\tilde{\omega}\psi_{l}^{f1\omega}+\frac{M}{b}\psi_{l}^{f1\phi}+k_z\psi_{l}^{f1z}\,,
\ee 
where $M/b$ is small. The leading order equations are  
\be
\Bigg(\Gamma^{\underline{r}}\partial_r
+(-1)^l\frac{m_f}{\sqrt{u}}\Bigg)\psi_{l}^{f0}=0\,.
\ee 
Before knowing how infalling boundary conditions from the near region result in the far region, we have in the far region eight independent horizon initial boundary conditions and linear order solutions could be determined from leading order ones. The eight initial boundary conditions for $\begin{pmatrix}\psi_1^{f0}\\ \psi_2^{f0}\end{pmatrix}$ are 
\be\label{8bcs}
V_\text{ini}^j=\left( r^{m_f}\delta_1^j\,,~r^{m_f}\delta_2^j\,,~r^{-m_f}\delta_3^j\,,~r^{-m_f}\delta_4^j\,,~r^{-m_f}\delta_5^j\,,~r^{-m_f}\delta_6^j\,,~r^{m_f}\delta_7^j\,,~r^{m_f}\delta_8^j\right)^T\,,~j\in\{1,...,8\}\,.\nn
\ee 
The leading order solutions are just exact solutions $r^{\pm m_f}$ for the components with nonzero boundary conditions. Note that later we will use the matching region solution to reduce these eight linearly independent boundary conditions to four by imposing infalling boundary conditions. 

The initial boundary conditions at the horizon for various components of $\psi_{l}^{f1\omega,z}$ are $r^{-(-1)^l m_f-1}$ as they are sourced by leading order solutions. Thus the first order corrections in $\omega$ and $k_z$ for far region solutions are not important as for pure $AdS$ background the solutions $r^{-(-1)^l m_f-1}$ are exact solutions and do not affect the boundary source and expectation terms.  
From here on we focus on $\psi_{l}^{f1\phi}$.

The initial boundary conditions for $\psi_{l}^{f1\phi}$ are all zero. We integrate the far region equations for $\psi_{l}^{f1\phi}$ and get the following result. Under the boundary condition $V_\text{ini}^{1,2}$, the only nonzero boundary values are $\psi_{2}^+=P_1 \frac{M}{b} r^{1/4} \begin{pmatrix} \delta_1^j\\ \delta_2^j \end{pmatrix}$; for $V_\text{ini}^{3,4}$, only nonzero boundary values are $\psi_{2}^-=-P_2 \frac{M}{b} r^{-1/4}\begin{pmatrix} \delta_3^j\\ \delta_4^j \end{pmatrix}$; for $V_\text{ini}^{5,6}$, only nonzero boundary values are $\psi_{1}^+=P_2 \frac{M}{b} r^{-1/4} \begin{pmatrix} \delta_5^j\\ \delta_6^j \end{pmatrix}$; for $V_\text{ini}^{7,8}$, only nonzero boundary values are $\psi_{1}^-=-P_1 \frac{M}{b} r^{1/4} \begin{pmatrix} \delta_7^j\\ \delta_8^j \end{pmatrix}$, where $P_{1,2}$ are numbers determined by numerics and for the set of parameters that we have used in this paper, we have $P_1\simeq 503.5$ and $P_2\simeq 0.0073$. %\comment{CHECK}

Now we impose the following four infalling boundary contions at the horizon of the near region with ${\begin{pmatrix}a_1^{+j}\\ a_2^{+j}\end{pmatrix}}_i=\delta_i^j,~i,j\in\{1...4\}$. 
At the matching region, expanding the solutions we get the near horizon boundary conditions for the far region solutions. Thus we have the following far region solutions for these four infalling boundary conditions after matching the coefficients in the matching region. For the first boundary condition, the far region solution with infalling boundary condition obtained from the matching region is 
 \be -\frac{i 2^{1/4}\Gamma[1/4]}{
\pi k_{1\omega}^{1/4}} \psi_{l}^{1}+\frac{(\omega+a_0)2^{1/4}}{k_{1\omega}^{7/4}}\left(\frac{\sqrt{2}}{\Gamma(1/4)}+\frac{i \Gamma(3/4)}{\pi}\right)\psi_{l}^2\,;
\ee
for the second boundary condition, the far region solution is
\be
-\frac{i 2^{1/4}\Gamma[1/4]}{
\pi k_{1\omega}^{1/4}} \psi_{l}^{3}+\frac{(\omega-a_0)2^{1/4}}{k_{1\omega}^{7/4}}\left(\frac{\sqrt{2}}{\Gamma(1/4)}+\frac{i \Gamma(3/4)}{\pi}\right)\psi_{l}^4\,;
\ee for the third boundary condition, the far region solution is
\be
-\frac{i 2^{3/4}\Gamma[3/4]}{
\pi k_{1\omega}^{3/4}} \psi_{l}^{5}+\frac{(\omega-a_0)2^{-3/4}}{k_{1\omega}^{5/4}}\left(\frac{2}{\Gamma(3/4)}-\frac{i\sqrt{2} \Gamma(1/4)}{\pi}\right)\psi_{l}^6\,;
\ee and for the fourth boundary condition, the far region solution is
\be
-\frac{i 2^{3/4}\Gamma[3/4]}{
\pi k_{1\omega}^{3/4}} \psi_{l}^{7}+\frac{(\omega-a_0)2^{-3/4}}{k_{1\omega}^{5/4}}\left(\frac{2}{\Gamma(3/4)}-\frac{i\sqrt{2} \Gamma(1/4)}{\pi}\right)\psi_{l}^8\,,
\ee where $\psi_{l}^j$ corresponds to far region solutions under the $j$-th boundary condition (\ref{8bcs}). Here the value of $k_{1\omega}$ at $k_z=0$ is the same as $k_{2\omega}$ at $k_z=0$.

With these four linearly independent solutions and the boundary values of $\psi_{1,2}^j$ that we have already obtained earlier, we could now get the source and expectation matrices, which are very long and we do not write them out here. Now the determinant of the source matrix has an extra $P_2^2$ term compared to the $M/b=0$ case meaning that the determinant of the source matrix is not zero anymore at $\omega=a_0,~k_z=0$ due to nonzero $P_2^2$, which makes the $k_{1\omega}=0$ pole vanish and becomes a pseudogap.

%%%%%%%%%%%%%%%%%%%%%%%%%%%%%%%%%%%%%%%%%%%
\subsection{Topological invariants for holographic nodal line semimetal}
%%%%%%%%%%%%%%%%%%%%%%%%%%%%%%%%%%%%%%%%%%%

For a nodal line semimetal, there are two topological invariants as shown in \cite{rev1}. The first one is to take a circle linking the nodal loop in the momentum space of $k_x$, $k_y$ and $k_z$ and this loop cannot shrink to a point as it cannot deform adiabatically to unlink the nodal loop. The second topological invariant is defined on a sphere enclosing the nodal loop. In this case, the sphere also can not shrink to a point without passing singularities in the Green function. The first topological invariant is the one responsible for the stability of the nodal loop under small perturbations, i.e. the nodal line semimetal does not become gapped under small perturbations. The second topological invariant is related to whether the critical point is topological or not. As the second topological invariant would require too much numerics, we will not consider this one in this paper. 

\begin{figure}[h!]
\begin{center}
\includegraphics[width=0.24\textwidth]{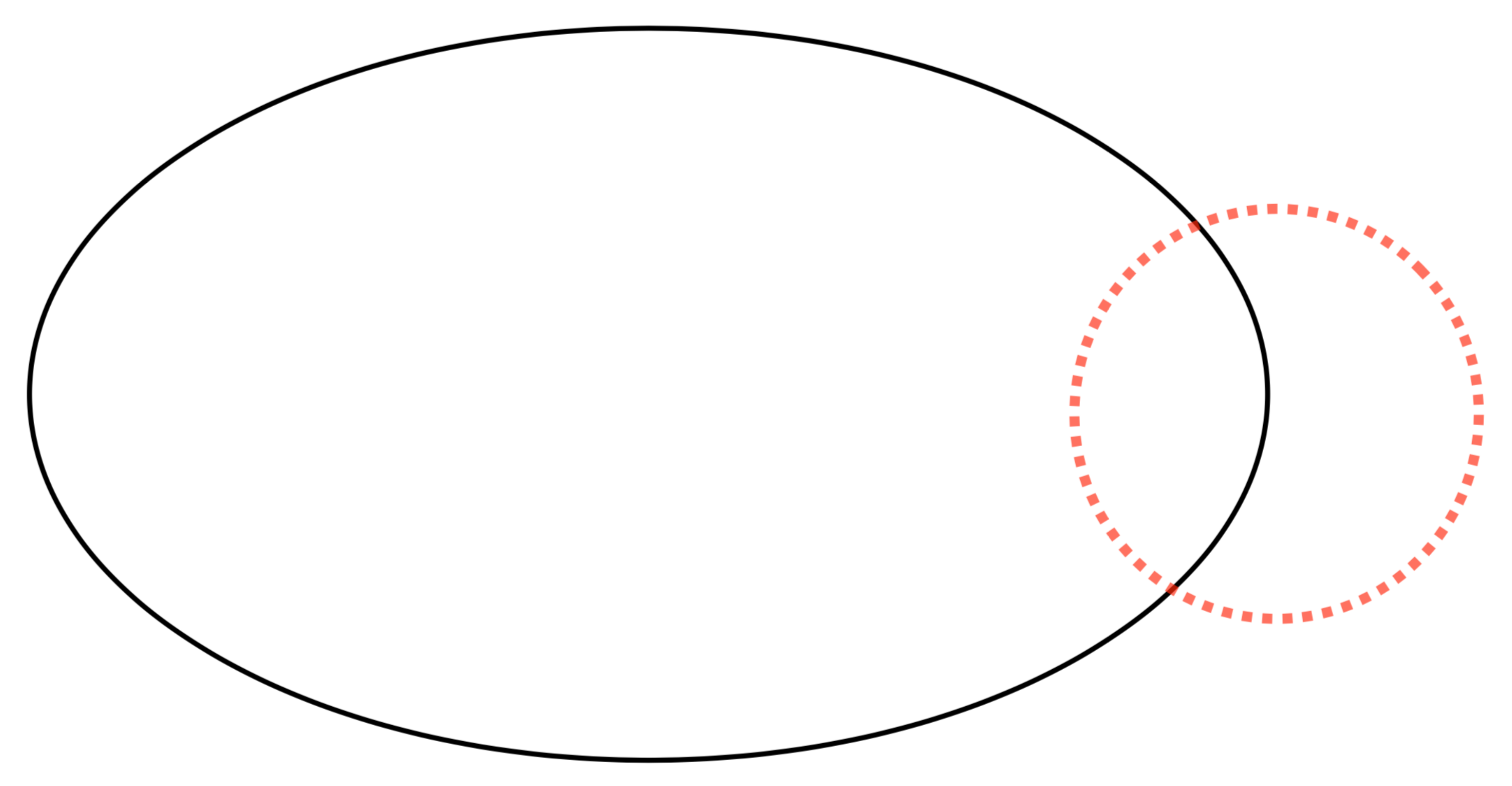}
\includegraphics[width=0.24\textwidth]{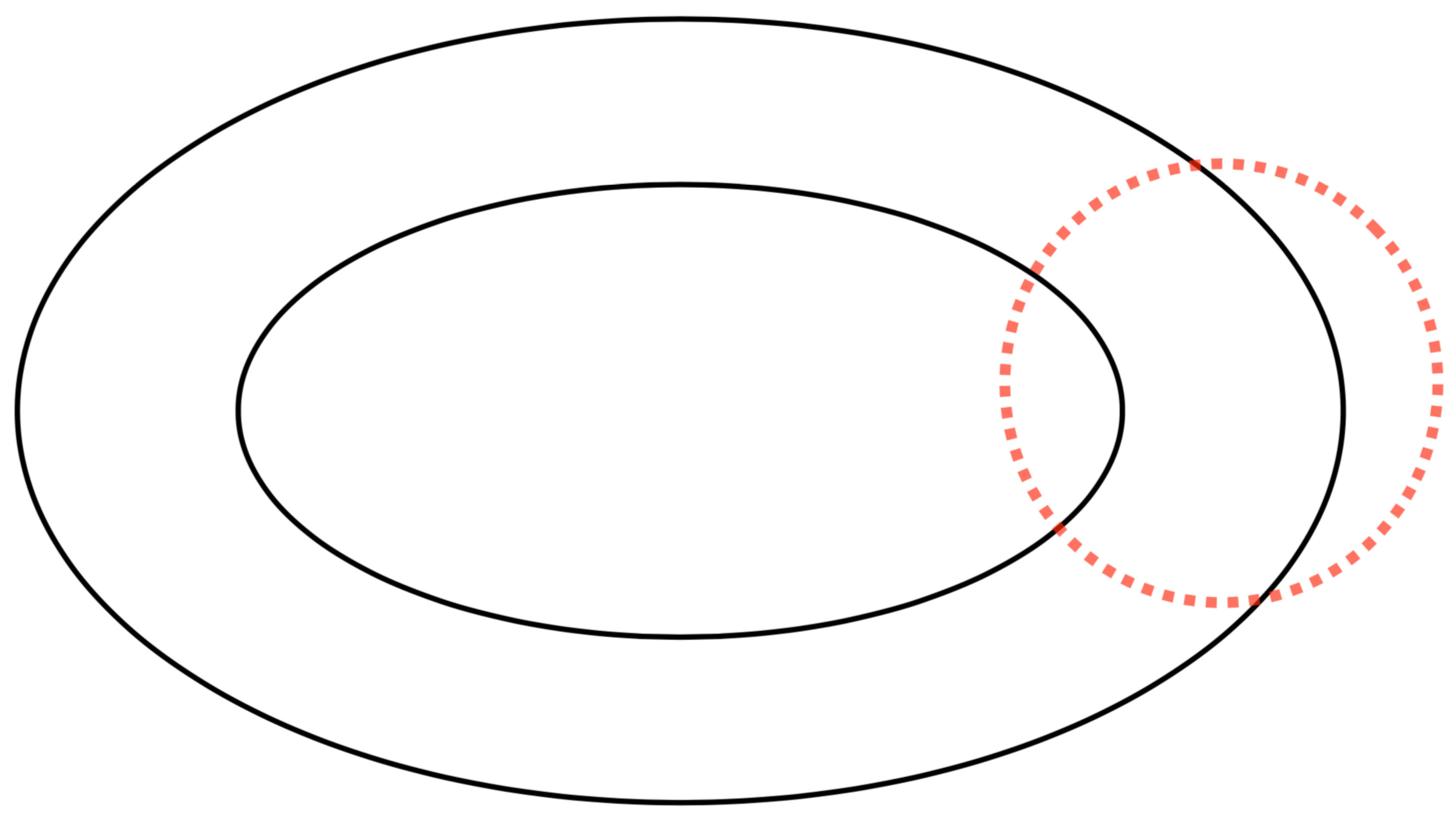}
\includegraphics[width=0.24\textwidth]{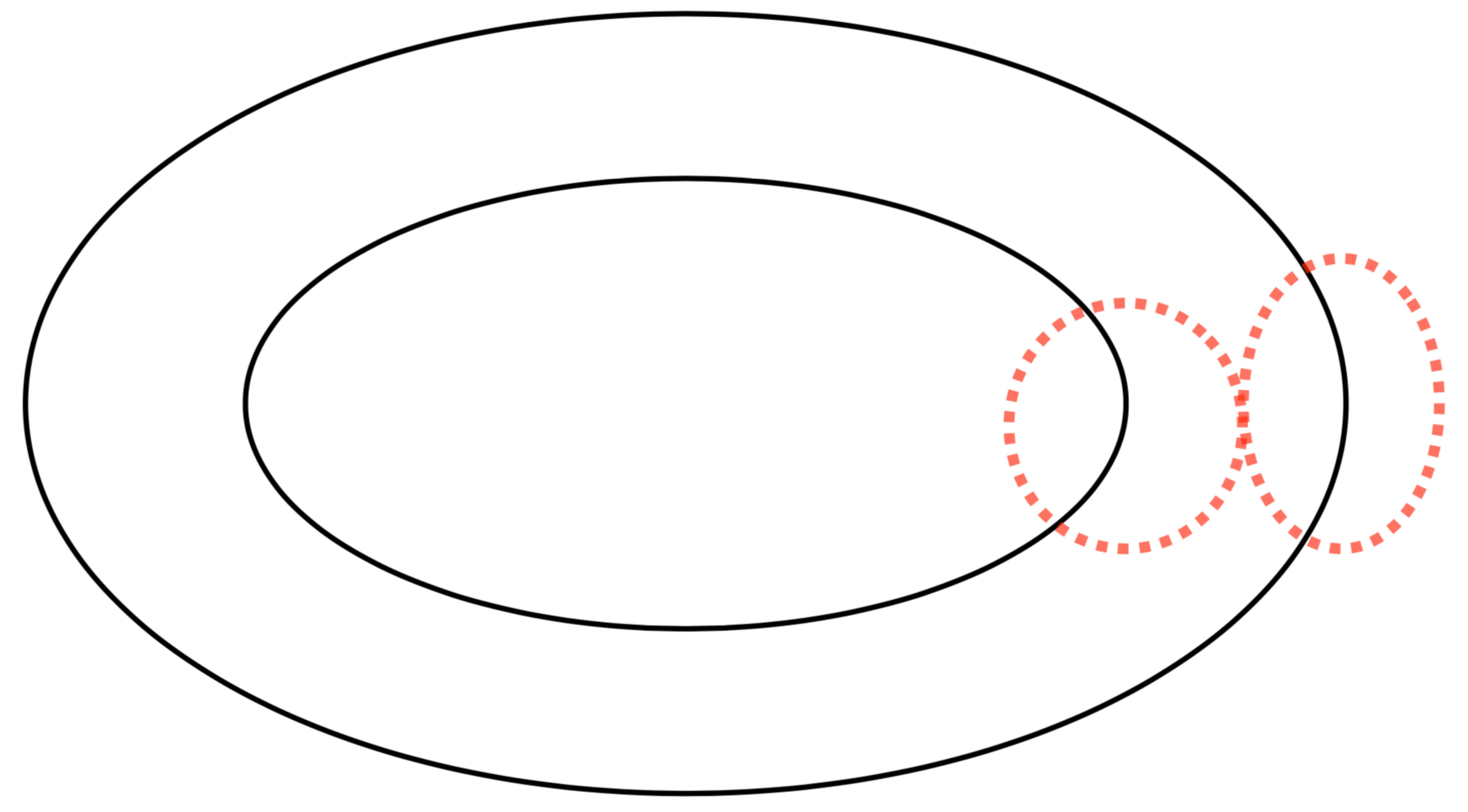}
\includegraphics[width=0.24\textwidth]{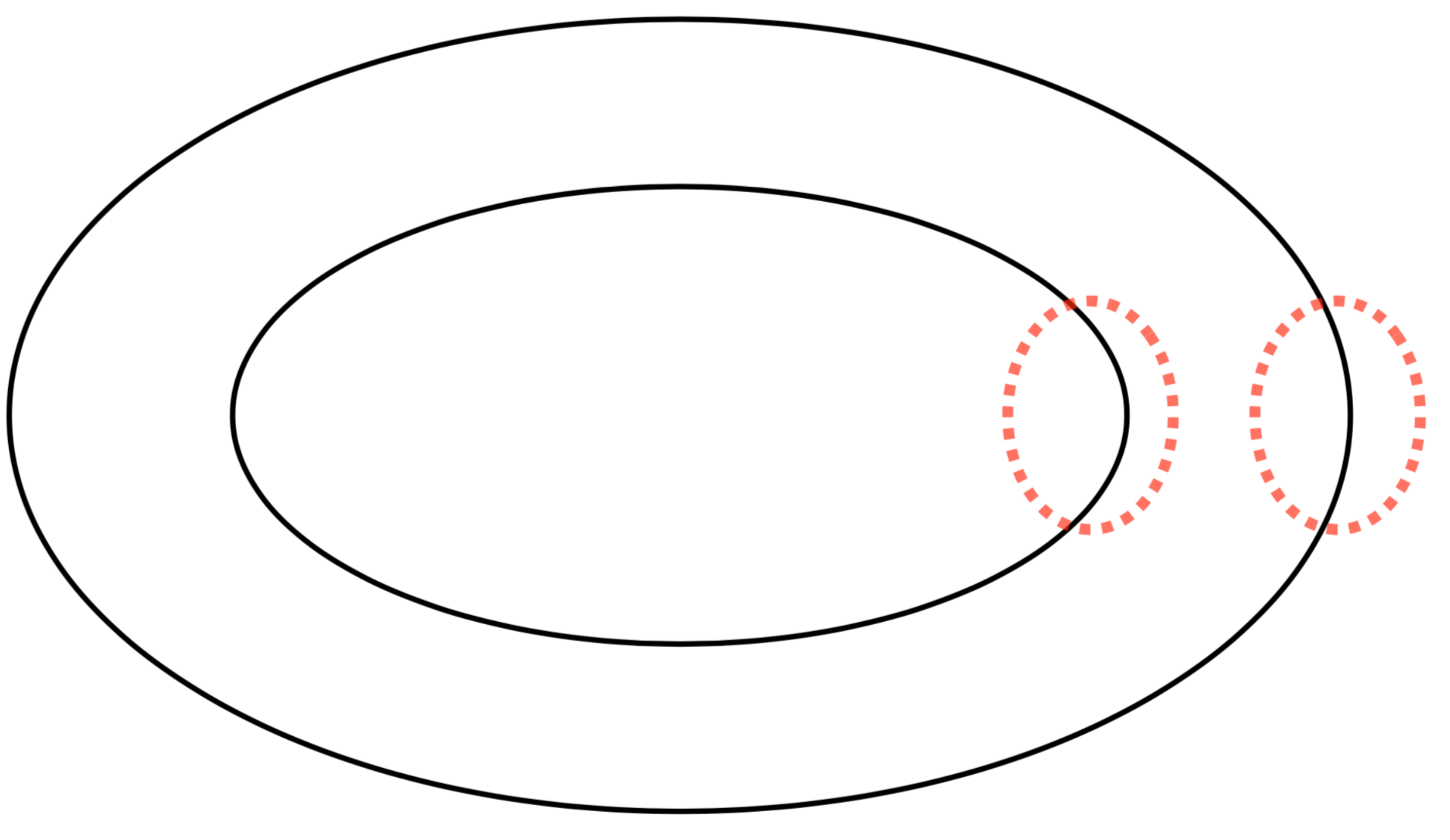}
\end{center}
\vspace{-0.4cm}
\caption{\small Illustration of the circle to which the Berry phase is associated. From left to right: a circle with only one nodal loop passing through its inside; a circle with two nodal loops passing through its inside; the circle could continuously deform to two separated circles, each of which is of the type in the first figure.}
\label{fig:berry}
\end{figure}

The first topological invariant is a Berry phase along the circle. We have multiple while discrete nodal lines in the $k_x$-$k_y$ plane at $k_z=0$ in the holographic nodal line semimetal phase and for each nodal line we could define a Berry phase. For each two or even more nodal lines we could also define a circle linking at the same time with two or more nodal lines, i.e. two or more nodal lines pass through the inside of the circle, which however could be continuously deformed to two or more separate circles of each nodal line itself as is shown in Fig. \ref{fig:berry}. Thus in the following we will focus on the Berry phase of each nodal line.

To avoid tedious numerical calculations, we choose very closely located discrete points on the loop and calculate the Berry phase in the discrete limit. We will show that in this case, the effective topological Hamiltonian method is still applicable and a nontrivial Berry phase of $\pi$ could be obtained for the holographic nodal line semimetal phase.

The procedure to calculate the Berry phase is the following. We first find the position of the Fermi surface $k_F=\sqrt{k_x^2+k_y^2}$ at $k_z=0$ and $\omega=0$. Then without loss of generality we take the circle in the $k_x$-$k_z$ plane to be $\sqrt{k_z^2+(k_x-k_F)^2}=k_0$ and $k_y=0$. Along this circle, we choose $N$ points to be $k_z=c_f \cos\theta$ and $k_x=k_F+c_f \sin\theta$ where  $\theta=\frac{2\pi j}{N}$, $j\in \{1,...N\}$ and the range of $\theta$ covers $0$ to $2\pi$ as shown in Fig. \ref{fig:npoints}. $c_f$ should be chosen to be small enough such that the circle does not pass through another nodal line. Then we could define the Berry phase using the discrete version and calculate the total Berry phase acquired along this circle. The discrete Berry phase is defined as
\be\label{berryphase}
e^{-i \phi_{i_1i_2}}=\frac{\langle n_{i_1}|n_{i_2}\rangle}{|\langle n_{i_1}|n_{i_2}\rangle|}\,,
\ee 
where $| n_{i_1}\rangle$ and $|n_{i_2}\rangle$ are two adjacent eigenstates along the circle. The total Berry phase is the sum of all adjacent phases along the circle from $0$ to $2\pi$. 

\begin{figure}[h!]
\begin{center}
\includegraphics[width=0.52\textwidth]{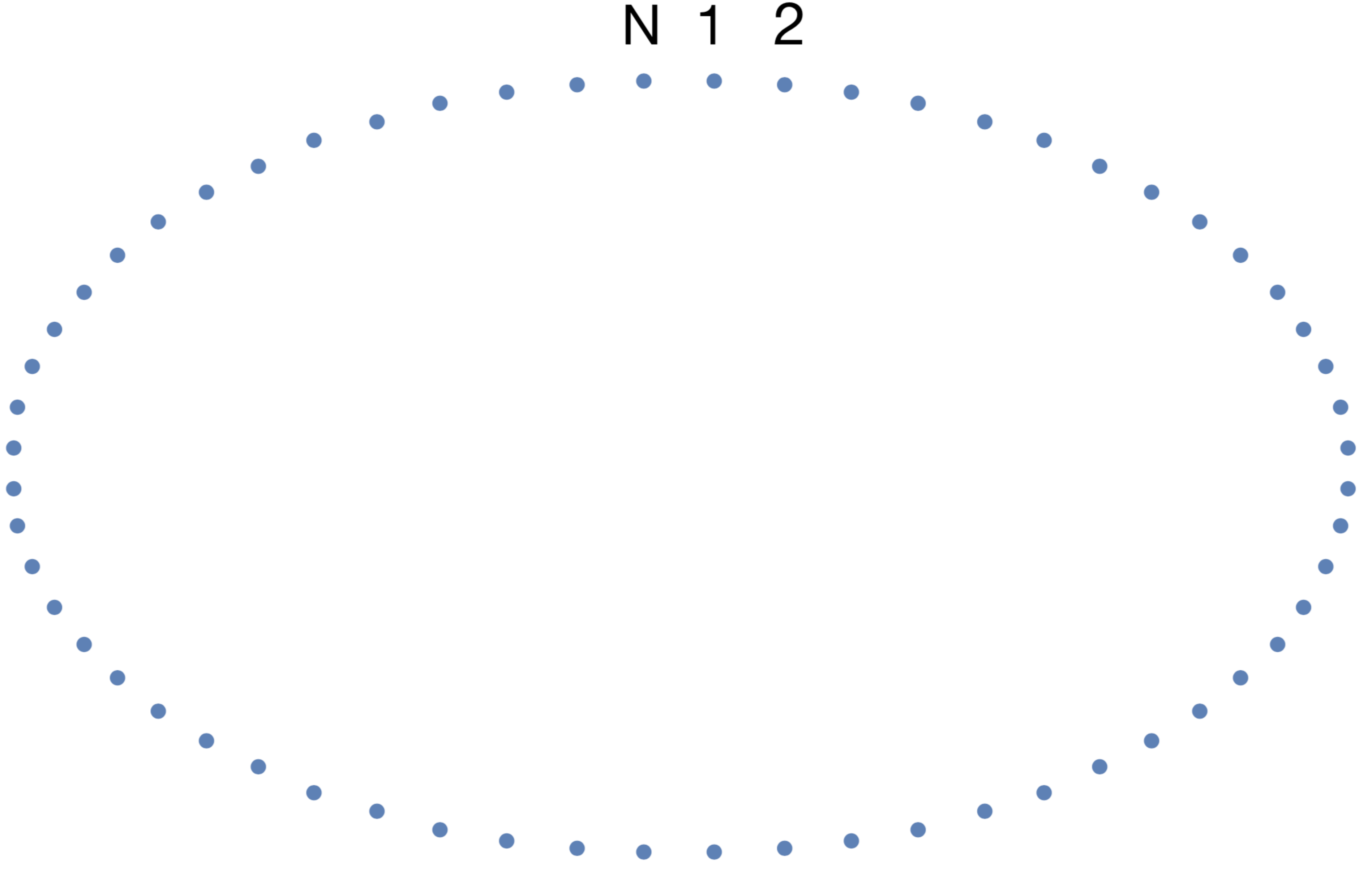}
\end{center}
\vspace{-0.4cm}
\caption{\small We can discrete the circle in the $k_x$-$k_z$ plane with $N$ points.}
\label{fig:npoints}
\end{figure}

For the nodal line semimetal, the poles are also band crossing points. At $\omega=0$ and $k_z\neq 0$, the near horizon boundary condition is proportional to $e^{-|k_z|/(u_0 r)}$ which is real. The four eigenvalues of $G^{-1}(0,{\bf k})$ are real and appear in $\pm$ pairs.  The eigenstates of $G^{-1}(0,{\bf k})$ are also real. This feature is the same as the weakly coupled theory for a nodal line semimetal and this means that the relative phase between adjacent eigenvectors could either be $0$ or $\pi$. For $k_z\to 0$ the near horizon boundary condition becomes tricky as the $k_z$ and $k_x$ contributions may be equally large and we cannot ignore $k_x$ terms anymore. Thus we first choose discrete points on the circle not very close to the $k_x$ axis. 

Using numerics, we choose 51 discrete points on a circle with $|k-k_F|=c_f$ where $c_f$ is a small number. To see more clearly whether there is a phase change on this circle, we have the following Fig. \ref{fig:berry-nlsm1} of the four components of the normalized gapless negative eigenvalued eigenvector for $k_F=931/1000$\footnote{Note that we have fixed $b=1$.} at $M/b\simeq 0.0013$ and the qualitative behavior for other nodal lines is the same. For the eigenvector at each site, there is a freedom to multiply the eigenvector by $\pm 1$ and whether we should choose $1$ or $-1$ is determined by the continuity of the vectors, i.e. we choose the eigenvector which continuously evolves on the circle. 
\begin{figure}[h!]
\begin{center}
\includegraphics[width=0.24\textwidth]{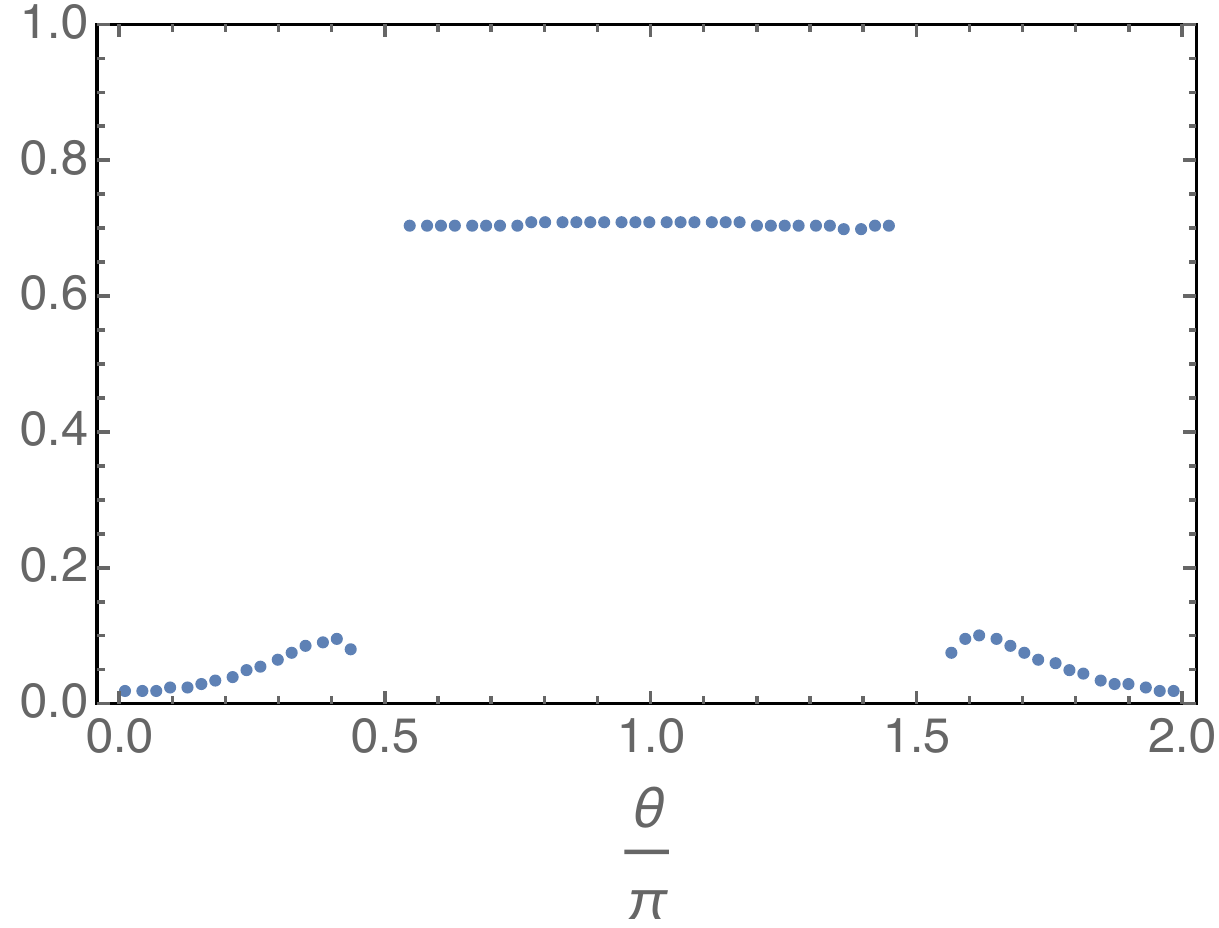}
\includegraphics[width=0.24\textwidth]{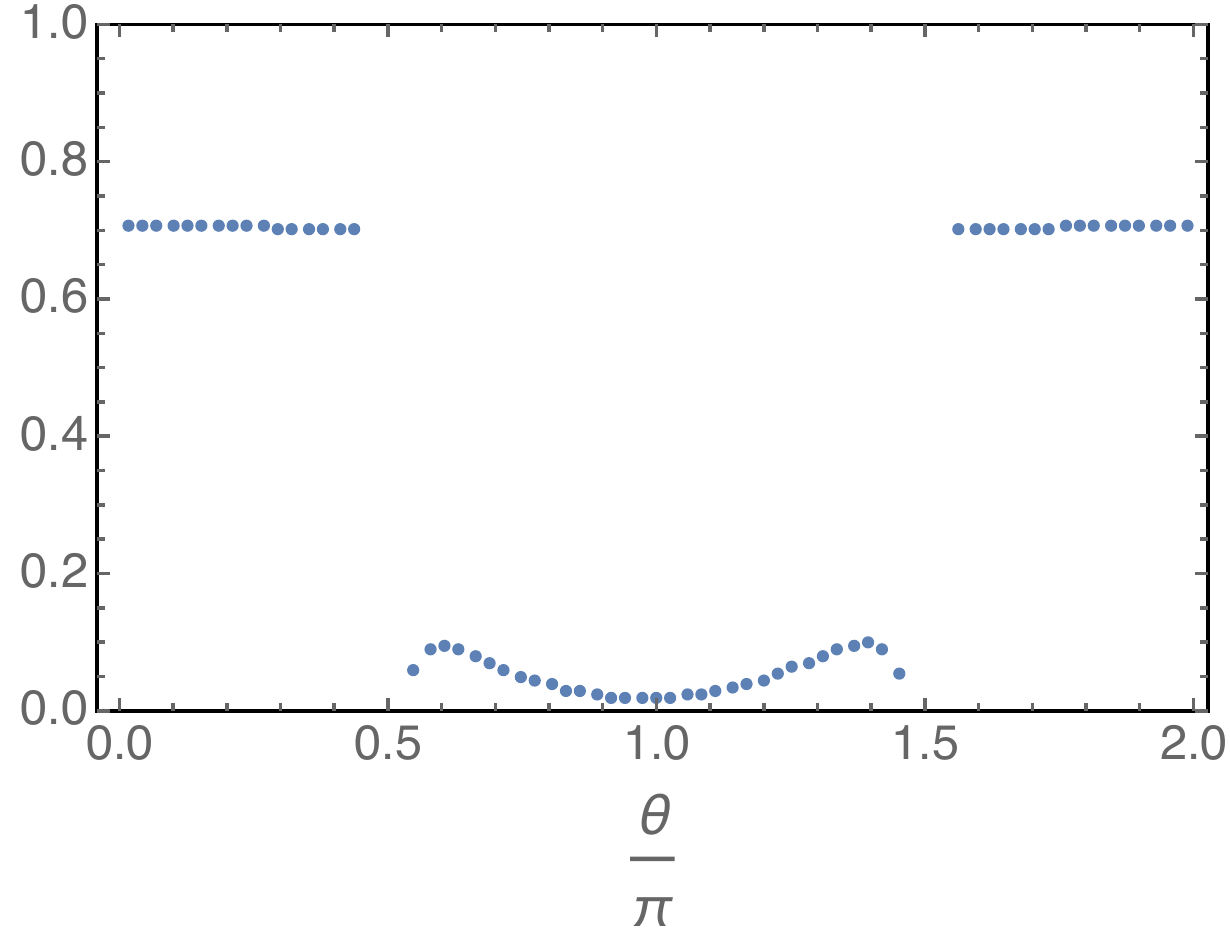}
\includegraphics[width=0.24\textwidth]{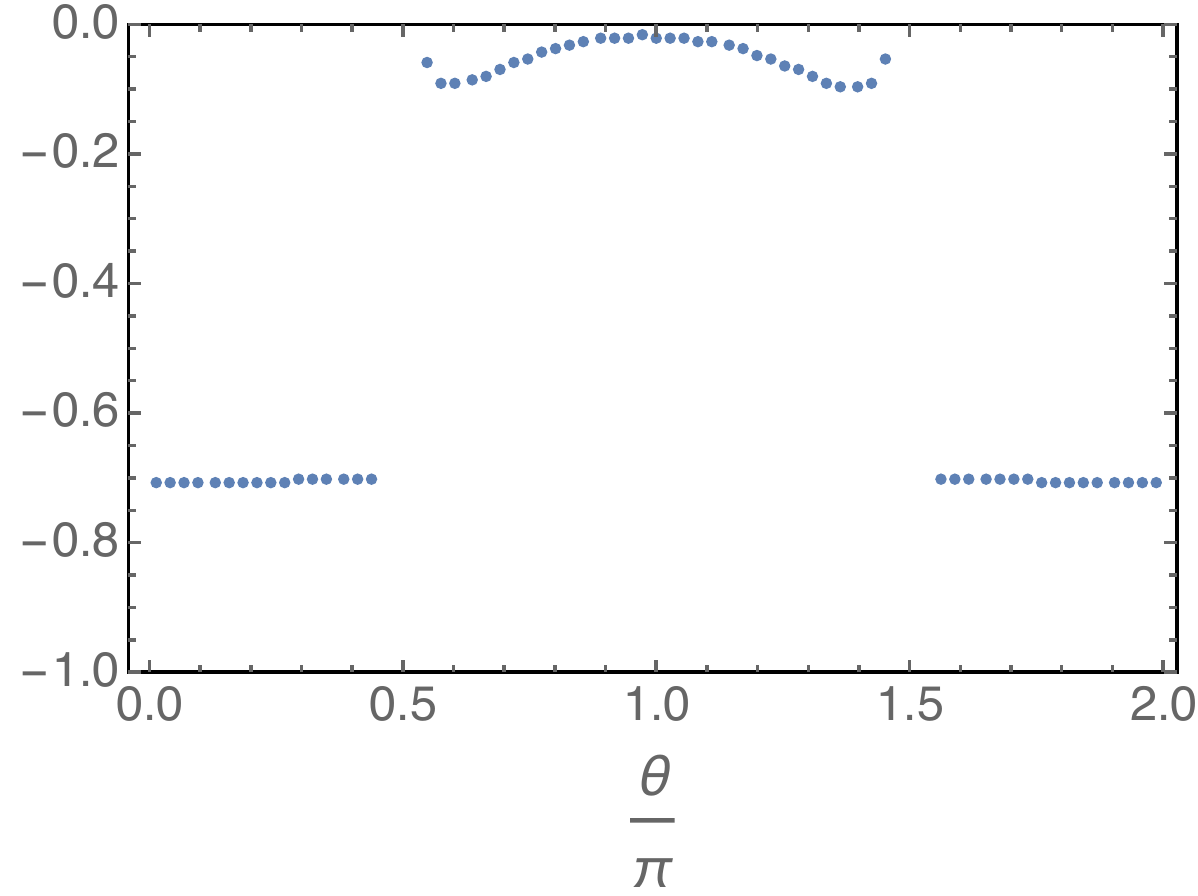}
\includegraphics[width=0.24\textwidth]{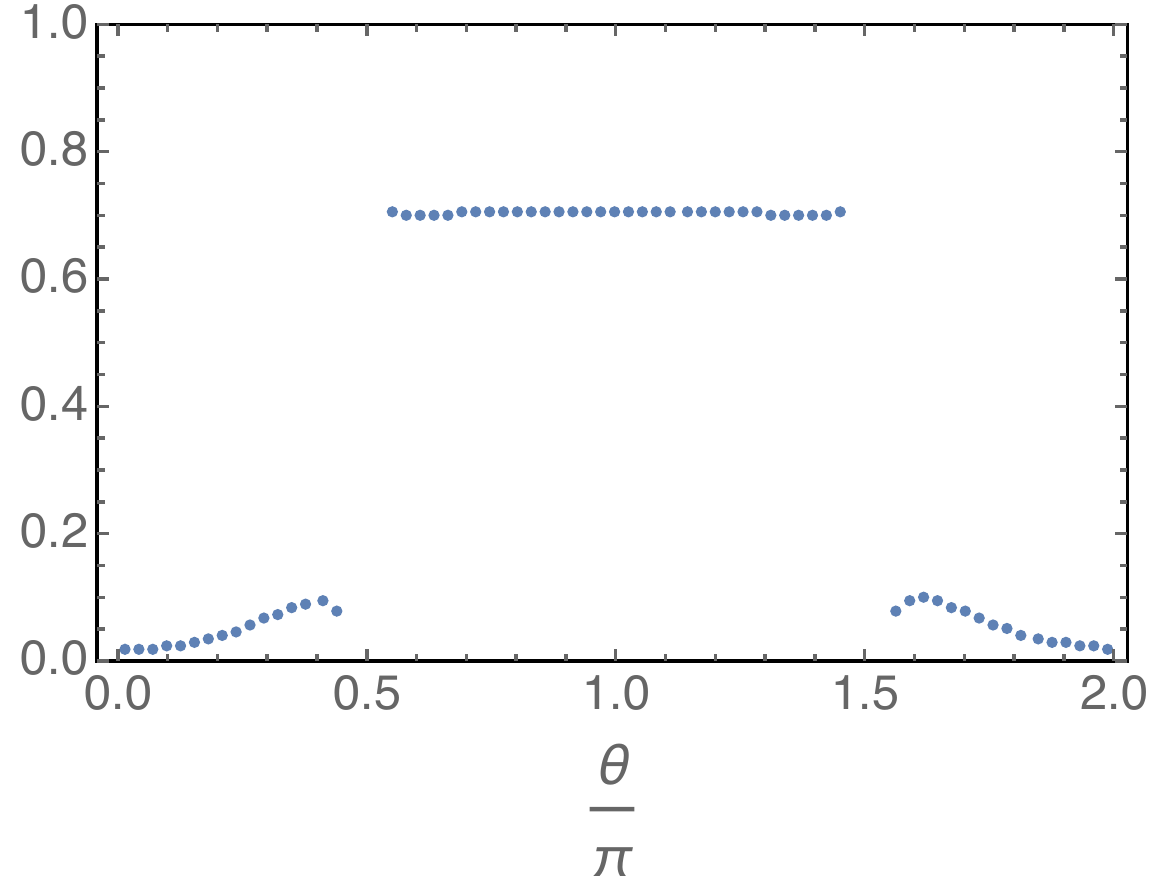}
\end{center}
\vspace{-0.7cm}
\caption{\small The value of the four components of the normalized gapless negative eigenvalued eigenvector of the topological Hamiltonian, i.e. $-G^{-1}(0,{\bf k})$ of the holographic nodal line semimetal at $M/b\simeq 0.0013$ along the circle around the pole $k_F\simeq 0.931$. $\theta=\pi/2, 3/2\pi$ is $k_z=0$. This behavior is qualitatively the same for other poles and for small deformations of the circle that does not pass through the nodal lines.}
\label{fig:berry-nlsm1}
\end{figure}

From this figure, we could see that in the $k_z>0$ and $k_z<0$ regions, the behavior of the eigenvectors are quite different. In the $k_z<0$ region, the first and the fourth components are equal and are close to the value $1/\sqrt{2}$ while the second and the third components are opposite %equal
and are very small, i.e. the eigenvectors approach $|n_{k_z\to 0_-}\rangle=(1/\sqrt{2},0,0,1/\sqrt{2})^T$ when $k_z\to 0$ from below. In the $k_z>0$ region, the second and third components are close to $\pm 1/\sqrt{2}$ while the first and the fourth components are almost zero, i.e. the eigenvectors approach $|n_{k_z\to 0_+}\rangle=(0,1/\sqrt{2},-1/\sqrt{2},0)^T$. This shows that at $k_z=0$ there is a sudden jump in the eigenvectors that the adjacent eigenvectors are orthogonal to each other, i.e. $\langle n_{k_z\to 0_-}|n_{k_z\to 0_+}\rangle=0$. According to the formula of discrete Berry phases, this gives undetermined Berry phases. However, in fact though $|n_{k_z\to 0_+}\rangle$ and $|n_{k_z\to 0_-}\rangle$ are orthogonal to each other, it could be that the eigenvector $n_{k_z=0}$ at $k_z=0$ is not orthogonal to either of  $|n_{k_z\to 0_+}\rangle$ and $|n_{k_z\to 0_-}\rangle$ and gives a determined result for the Berry phase. Thus the eigenvectors at the $k_z=0$ points play an important key role to determine the Berry phase. %\comment{CHECK}

\begin{figure}[h!]
\begin{center}
\includegraphics[width=0.67\textwidth]{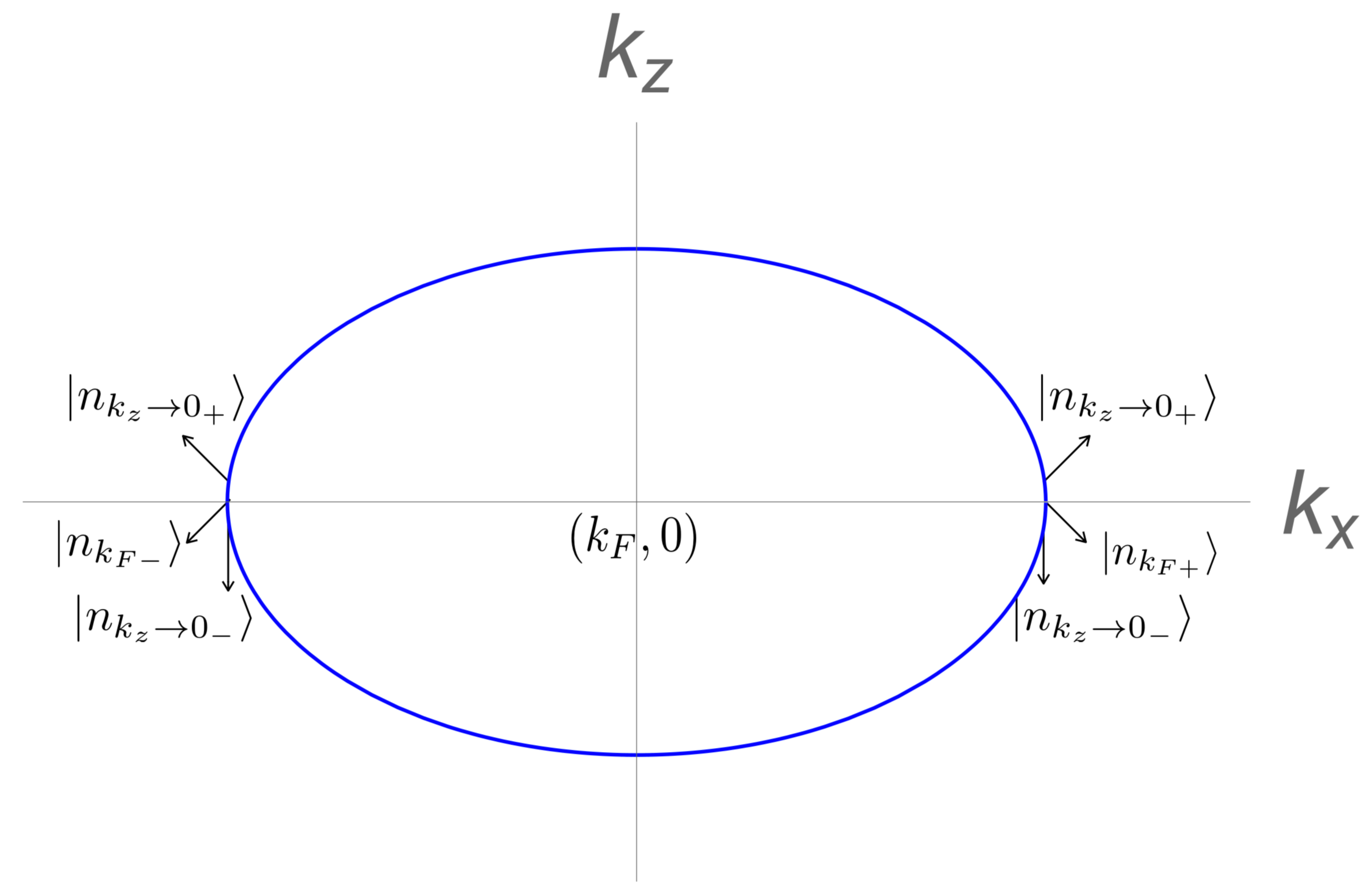}
\end{center}
\vspace{-0.6cm}
\caption{\small  Illustration for the calculation of Berry phase for holographic nodal line semimetal phase around each pole $k_{F,i}$ or any zero point $k_{0,i}$ of the Green function. $| n_{k_z\to 0_\pm}\rangle$ are the same for all these points while $|n_{k_{F\pm}}\rangle$ are different depending on the points. }
\label{fig:berry-illus}
\end{figure}

Numerics could not detect small but nonzero $k_z$ regions very accurately, but we could work directly at $k_z=0$ which is easier in numerics. For each of the pole, the small circle would intersect with the $k_x$ axis twice (i.e. $k_z=0$) one at $k_{F-}=k_F(1-\delta)$ and one at $k_{F+}=k_F(1+\delta)$, where $\delta\ll 1$ is a small number. We could work out the negative valued eigenvector of both $k_{F-}$ and $k_{F+}$. We find that for all the poles from the same two bands as $k_F\simeq 1.048$, the eigenvector at $k_{F-}$ is $|n_{k_{F-}}\rangle=1/2(1,1,-1,1)^T$ while the eigenvector at $k_{F+}$ is $|n_{k_{F+}}\rangle=1/2(1,-1,1,1)^T$. To connect these two eigenvectors with those of $| n_{k_z\to 0_-}\rangle$ and $ |n_{k_z\to 0_+}\rangle$, we find that there needs to be a $\pi$ phase along the circle. When we first connect $ |n_{k_z\to 0_-}\rangle$ and $| n_{k_z\to 0_+}\rangle$ to $|n_{k_{F-}}\rangle$ we find that  $| n_{k_z\to 0_-}\rangle$ and $| n_{k_z\to 0_+}\rangle$ written in this way are already continuously connected without flipping signs of either of the two vectors. When we connect $ |n_{k_z\to 0_-}\rangle$ and $| n_{k_z\to 0_+}\rangle$ to $|n_{k_{F+}}\rangle$ we find that either one of $| n_{k_z\to 0_-}\rangle$ and $ |n_{k_z\to 0_+}\rangle$ has to flip the sign or there would be a $\pi$ phase change at $k_z=0$ and if we flip the sign of one of $| n_{k_z\to 0_-}\rangle$ and $| n_{k_z\to 0_+}\rangle$, a $\pi$ phase difference would appear in the upper or lower half plane in the $k_x$-$k_z$ plane. Thus for these poles, we could see that there is a nontrivial Berry phase of $\pi$. An illustration on the vectors can be found in Fig. \ref{fig:berry-illus} and the different vectors for different poles or zeros can be found in Tab. \ref{tab:vector}.  

\begin{table}[h!]
\begin{center}
\setlength{\tabcolsep}{6pt}
\renewcommand{\arraystretch}{1.4}
\begin{tabular}{ |c||c|c|c|  }
 \hline
 & poles from Bands I & poles from Bands II & zeros of the Green function\\
 \hline
  \hline
 $| n_{k_{F+}}\rangle$ &$1/2(1,-1,1,1)^T$ & $1/2(1,-1,-1,1)^T$ &  $1/2(1,1,-1,1)^T$\\
 $| n_{k_{F-}}\rangle$    &$1/2(1,1,-1,1)^T$ & $1/2(1,1,1,-1)^T $ &  $1/2(1,1,-1,1)^T$\\
 \hline
   $| n_{k_z\to 0_+}\rangle$ &   \multicolumn{3}{|c|}{$(0,1/\sqrt{2},-1/\sqrt{2},0)^T$}  \\
 $| n_{k_z\to 0_-}\rangle$ &  \multicolumn{3}{|c|}{$(1/\sqrt{2},0,0,1/\sqrt{2})^T$}   \\
   \hline
\end{tabular}
\end{center}
\vspace{-0.4cm}
\caption{\small A table of $|n_{k_{F\pm}}\rangle$ and $| n_{k_z\to 0_\pm}\rangle$ for poles from bands I, II and zeros of the Green function.}
\label{tab:vector}
\end{table}

The behavior of the negative valued eigenvectors for $k_z\neq 0$ points on the small circle around the pole is the same for all the poles and all the zeros of the Green functions, while the $k_z=0$ and $k_{F,\pm}=k_F(1\pm\delta)$ negative valued eigenvectors 
$|n_{k_{F\pm}} \rangle$ are different depending on whether the poles come from bands I or II. In general, for poles from bands I, i.e. the blue colored bands in Fig. \ref{fig:spec}, $|n_{k_{F\pm}}\rangle$ are the same as above and all result in a nontrivial Berry phase of $\pi$. For the poles from bands II, e.g. for $k_F=931/1000$, the eigenvector at $k_{F-}$ is $|n_{k_{F-}}\rangle=1/2(1,1,1,-1)^T$ while the eigenvector at $k_{F+}$ is $|n_{k_{F+}}\rangle=1/2(1,-1,-1,1)^T$, which are still orthogonal to both  $| n_{k_z\to 0_-}\rangle$ and $| n_{k_z\to 0_+}\rangle$. This means that for the poles from the bands II, the Berry phase for the circle around the nodal point is still undetermined. 

Besides checking the Berry phase for the poles, we have also checked if there is a nontrivial Berry phase at each zero of the Green functions $G(0,k_{x}^{i})=0$ and we find that for the zeros of the Green functions, $|n_{k_{F+}}\rangle=|n_{k_{F-}}\rangle=1/2(1,1,-1,1)^T$ and results in a trivial Berry phase of 0. This is different from the Weyl semimetal case where the zeros of the Green function could still have nontrivial topological invariants and this may also indicate that for positive $m_f$ the poles do not have nontrivial Berry phases.

Thus with the above we conclude that for the holographic nodal line semimetal phase, there is a nontrivial topological invariant associated with poles from bands I and for poles from bands II the Berry phase is undetermined.

%%%%%%%%%%%%%%%%%%%%%%%%%%%%%
\section{Conclusion and discussion}
\label{sec5}
%%%%%%%%%%%%%%%%%%%%%%%%%%%%

We have calculated the topological invariants for holographic Weyl and nodal line semimetals. For both cases, we find that we could define a nontrivial topological invariant using the topological Hamiltonian method, which allows us to calculate the topological invariants using the zero frequency Green functions of fermionic operators. For the holographic WSM case, semi-analytic calculations allows us to get the topological invariants for very small $M/b$, which are $\pm 1$ and are exactly the same as weakly coupled WSM model, while for larger $M/b$ we will have to use numerics. For the holographic NLSM case, different from the weakly coupled models, there are multiple nodal lines which are poles at $\omega=0$ with $k_{F,i}$ for the holographic model. From the zero frequency Green function we could tell that these poles come from different sets of bands indicating that the two gapped bands and two gapless bands exchange their roles alternatively along the $k_x$ axis. A discrete version of Berry phase calculation shows that for half of these poles there is a nontrivial $\pi$ Berry phase while for the other half coming from the other two bands, the Berry phase is undetermined.  

These nontrivial topological invariants provide a further robust %nontrivial %the last piece of 
evidence that the holographic models are strongly coupled topologically nontrivial semimetals and these holographic models serve as a useful arena and a useful tool for the study of various interesting properties of strongly topological semimetals. %The next step is also 
It would be interesting 
to generalize these to gapped systems and provide predictions of properties of strongly coupled gapped and gapless topological states of matter. 

%%%%%%%%%%%%%%%%%%%%%%%%%%%%%%%%%%%%%%%%%

%%%%%%%%%%%%%%%%%%%%%%%%%%%%%%%%%%%%%%%%%
\subsection*{Acknowledgments}
%%%%%%%%%%%%%%%%%%%%%%%%%%%%%%%%%%%%%%%%%
We would like to thank Rong-Gen Cai, Chen Fang, Carlos Hoyos, Elias Kiritsis, Karl Landsteiner,  Shun-Qing Shen, Sang-Jin Sin, Zhong Wang, Jan Zaanen, Long Zhang for useful discussions. This work is supported by the National Key R\&D Program of China (Grant No. 2018FYA0305800) and by the Thousand Young Talents Program of China. The work of Y.L. was also supported by the NFSC Grant No.11875083 and a grant from Beihang University. The work of Y.W.S. has also been partly supported by starting grants from UCAS and CAS, and by the Key Research Program of the Chinese Academy of Sciences (Grant No. XDPB08-1), the Strategic Priority Research Program of Chinese Academy of Sciences, Grant No. XDB28000000. We are also grateful to the hospitality of Hanyang University during the conference ``Holography and Geometry of Quantum Entanglement" (APCTP) where this work was presented.

%%%%%%%%%%%%%%%%%%%%%%%%%%%%%
\appendix
\section{$s$ and $e$}
\label{appA}
%%%%%%%%%%%%%%%%%%%%%%%%%%%%%

In this appendix we list the elements $s$ and $e$ appeared in Sec. \ref{secinapp}. Note that $x_i$ with $i\in 1, ..., 36$ are constants which are the boundary values of the solutions associated with the six far region boundary conditions.
\bea
s^1&=&\left( x_1+x_2 (k_z-a_0)\,,~-x_3 (k_x+i {k_y})\,,~x_4+x_5 (k_z-a_0)\,,~x_6 (k_x+i {k_y})\right)^T\nonumber\\
s^2&=&\left( x_3 (k_x-i {k_y})\,,~ x_1 + x_2  (k_z-a_0)\,,~ x_6(k_x-i {k_y})\,,~ -x_4 - x_5 (k_z-a_0)\right)^T\nonumber\\
s^3&=&\left(-x_{13} - x_{14} (k_z-a_0)\,,~-x_{15} (k_x + i k_y)\,,~ -x_{16} - x_{17}(k_z-a_0)\,,~ x_{18} (k_x + i k_y)\right)^T\nonumber\\
s^4&=&\left(-x_{15} (k_x - i k_y)\,,~ x_{13} + x_{14} (k_z-a_0)\,,~ -x_{18} (k_x - i k_y)\,,~ -x_{16} -x_{17} (k_z-a_0)\right)^T\nonumber\\
s^5&=&\left(x_{25} - x_{26} (k_z-a_0)\,,~ x_{27} (k_x+i {k_y})\,,~ x_{28} + x_{29} k_z\,,~ -x_{30} (k_x+i {k_y})\right)^T\nonumber\\
s^6&=&\left( x_{27} (k_x - i k_y)\,,~ -x_{25} + x_{26} (k_z-a_0)\,,~ x_{30} (k_x - i k_y)\,,~ x_{28} + x_{29} (k_z-a_0)\right)^T\nonumber
\eea
\bea
e^1&=&\left(-x_7 - x_8 k_z\,,~ x_9 (k_x+i {k_y})\,,~ -x_{10} - x_{11} (k_z-a_0)\,,~ -x_{12} (k_x+i {k_y})\right)^T\nonumber\\
e^2&=&\left(-x_9 (k_x - i k_y)\,,~ -x_7 - x_8 (k_z-a_0)\,,~ -x_{12} (k_x - i k_y)\,,~ x_{10} + x_{11} (k_z-a_0)\right)^T\nonumber\\
e^3&=&\left(x_{19} + x_{20} (k_z-a_0)\,,~ x_{21} (k_x+i {k_y})\,,~ x_{22} + x_{23} (k_z-a_0)\,,~ -x_{24} (k_x+i {k_y})\right)^T\nonumber\\
e^4&=&\left(x_{21} (k_x - i k_y)\,,~ -x_{19} - x_{20} k_z\,,~ x_{24} (k_x - i k_y)\,,~ x_{22} + x_{23} (k_z-a_0)\right)^T\nonumber\\
e^5&=&\left(-x_{31} - x_{32} (k_z-a_0)\,,~ -x_{33}(k_x+i {k_y})\,,~ -x_{34} + x_{35} (k_z-a_0)\,,~ x_{36} (k_x+i {k_y})\right)^T\nonumber\\
e^6&=&\left(-x_{33}(k_x - i k_y)\,,~ x_{31} + x_{32} (k_z-a_0)\,,~ -x_{36} (k_x - i k_y)\,,~ -x_{34} + x_{35} (k_z-a_0)\right)^T\nonumber
\eea

%\section{Appendix}

%In this appendix we will list the details 

%\bibliography{charge-diss}{}

\begin{thebibliography}{99}
% %%%%%%%%%%%%%%%%%%%%%%%%%
% %%%%%%%%%%%%%%%%%%%%%%%%%

%\cite{Witten:2015aoa}
\bibitem{Witten:2015aoa} 
  E.~Witten,
{\em Three Lectures On Topological Phases Of Matter,}
 % Riv.\ Nuovo Cim.\  {\bf 39}, no. 7, 313 (2016)
  \doi{10.1393/ncr/i2016-10125-3}{Riv.\ Nuovo Cim.\  {\bf 39}, no. 7, 313 (2016)} 
 [\arXiv{1510.07698}{cond-mat.mes-hall}].


 %\cite{Landsteiner:2015lsa}
\bibitem{Landsteiner:2015lsa} 
  K.~Landsteiner and Y.~Liu,
{\em The holographic Weyl semi-metal,}
 \doi{10.1016/j.physletb.2015.12.052}{Phys.\ Lett.\ B {\bf 753}, 453 (2016)}
  [\arXiv{1505.04772}{hep-th}].

%\cite{Landsteiner:2015pdh}
\bibitem{Landsteiner:2015pdh} 
  K.~Landsteiner, Y.~Liu and Y.~W.~Sun,
{\em Quantum phase transition between a topological and a trivial semimetal from holography,}
 \doi{10.1103/PhysRevLett.116.081602}{Phys.\ Rev.\ Lett.\  {\bf 116}, no. 8, 081602 (2016)}
  [\arXiv{1511.05505}{hep-th}].
  %%CITATION = doi:10.1103/PhysRevLett.116.081602;%%
  %15 citations counted in INSPIRE as of 16 Jun 2018

  %\cite{Liu:2018bye}
\bibitem{Liu:2018bye} 
  Y.~Liu and Y.~W.~Sun,
{\em Topological nodal line semimetals in holography,}
  \arXiv{1801.09357}{hep-th}.
  %%CITATION = ARXIV:1801.09357;%%
  %2 citations counted in INSPIRE as of 16 Jun 2018

%\cite{Zaanen:2015oix}
\bibitem{Zaanen:2015oix} 
  J.~Zaanen, Y.~W.~Sun, Y.~Liu and K.~Schalm,
 \href{http://www.cambridge.org/de/academic/subjects/physics/condensed-matter-physics-nanoscience-and-mesoscopic-physics/holographic-duality-condensed-matter-physics?format=HB#AlwhgydkVTSFfv7H.97}{\em Holographic Duality in Condensed Matter Physics,}  Cambridge University Press, 2015. 
  %%CITATION = INSPIRE-1384852;%%
  \bibitem{book0} 
M.~Ammon and J.~Erdmenger, 
\href{http://www.cambridge.org/de/academic/subjects/physics/theoretical-physics-and-mathematical-physics/gaugegravity-duality-foundations-and-applications#xOzmEecLSr4ZJFIH.97}{\em Gauge/gravity duality: Foundations and applications},  
Cambridge University Press, 2015.
%\cite{Hartnoll:2016apf}
\bibitem{review} 
  S.~A.~Hartnoll, A.~Lucas and S.~Sachdev,
{\em Holographic quantum matter,}
[\arXiv{1612.07324}{hep-th}].
  %%CITATION = ARXIV:1612.07324;%%



%\cite{Ammon:2016mwa}
\bibitem{Ammon:2016mwa} 
  M.~Ammon, M.~Heinrich, A.~Jimenez-Alba and S.~Moeckel,
{\em Surface States in Holographic Weyl Semimetals,}
  \doi{10.1103/PhysRevLett.118.201601}{Phys.\ Rev.\ Lett.\  {\bf 118}, no. 20, 201601 (2017)}
  [\arXiv{1612.00836}{hep-th}].
  %%CITATION = doi:10.1103/PhysRevLett.118.201601;%%   

  

%\cite{Landsteiner:2016stv}
\bibitem{Landsteiner:2016stv} 
  K.~Landsteiner, Y.~Liu and Y.~W.~Sun,
{\em Odd viscosity in the quantum critical region of a holographic Weyl semimetal,}
  \doi{10.1103/PhysRevLett.117.081604}{Phys.\ Rev.\ Lett.\  {\bf 117}, no. 8, 081604 (2016)}
  [\arXiv{1604.01346}{hep-th}].  

%\cite{Copetti:2016ewq}
\bibitem{Copetti:2016ewq} 
  C.~Copetti, J.~Fernandez-Pendas and K.~Landsteiner,
{\em Axial Hall effect and universality of holographic Weyl semi-metals,}
  %JHEP {\bf 1702}, 138 (2017)
  \doi{10.1007/JHEP02(2017)138}{JHEP {\bf 1702}, 138 (2017)}
  [\arXiv{1611.08125}{hep-th}].
  %%CITATION = doi:10.1007/JHEP02(2017)138;%%
  

  

  
  %\cite{Grignani:2016wyz}
\bibitem{Grignani:2016wyz} 
  G.~Grignani, A.~Marini, F.~Pena-Benitez and S.~Speziali,
{\em AC conductivity for a holographic Weyl Semimetal,}
  \doi{10.1007/JHEP03(2017)125}{ JHEP {\bf 1703}, 125 (2017)}
  [\arXiv{1612.00486}{cond-mat.str-el}].
  %%CITATION = doi:10.1007/JHEP03(2017)125;%%
  
 %\cite{Ammon:2018wzb}
\bibitem{Ammon:2018wzb} 
  M.~Ammon, M.~Baggioli, A.~Jiménez-Alba and S.~Moeckel,
{\em A smeared quantum phase transition in disordered holography,}
 % JHEP {\bf 1804}, 068 (2018)
 \doi{10.1007/JHEP04(2018)068}{JHEP {\bf 1804}, 068 (2018)}
  [\arXiv{1802.08650}{hep-th}].
  %%CITATION = doi:10.1007/JHEP04(2018)068;%%

 %\cite{Baggioli:2018afg}
\bibitem{Baggioli:2018afg} 
  M.~Baggioli, B.~Padhi, P.~W.~Phillips and C.~Setty,
{\em Conjecture on the Butterfly Velocity across a Quantum Phase Transition,}
% JHEP {\bf 1807}, 049 (2018)
 \doi{10.1007/JHEP07(2018)049}{JHEP {\bf 1807}, 049 (2018)} 
  [\arXiv{1805.01470}{hep-th}].
  %%CITATION = ARXIV:1805.01470;%%
 
%\cite{Gursoy:2012ie}
\bibitem{Gursoy:2012ie} 
  U.~Gursoy, V.~Jacobs, E.~Plauschinn, H.~Stoof and S.~Vandoren,
{\em Holographic models for undoped Weyl semimetals,} 
 % JHEP {\bf 1304}, 127 (2013)
  \doi{10.1007/JHEP04(2013)127}{JHEP {\bf 1304}, 127 (2013)}
  [\arXiv{1209.2593}{hep-th}].
  %%CITATION = doi:10.1007/JHEP04(2013)127;%%
  
%\cite{Hashimoto:2016ize}
\bibitem{Hashimoto:2016ize} 
  K.~Hashimoto, S.~Kinoshita, K.~Murata and T.~Oka,
{\em Holographic Floquet states I: a strongly coupled Weyl semimetal,}
 % JHEP {\bf 1705}, 127 (2017)
  \doi{10.1007/JHEP05(2017)127}{JHEP {\bf 1705}, 127 (2017)}
  [\arXiv{1611.03702}{hep-th}].
  %%CITATION = doi:10.1007/JHEP05(2017)127;%% 
 
    
 %\cite{Wang:2013ypa}
\bibitem{wang-prx} 
  Z.~Wang and S.~C.~Zhang,
{\em Topological Invariants and Ground-State Wave Functions of Topological Insulators on a Torus,}
\doi{10.1103/PhysRevX.4.011006}{Phys.\ Rev.\ X {\bf 4}, no. 1, 011006 (2014)}  
[\arXiv{1308.4900}{cond-mat.str-el}];  Z.~Wang and S.~C.~Zhang, {\em Simplified topological invariants for interacting insulators,}
\doi{10.1103/PhysRevX.2.031008}{Phys.\ Rev.\ X {\bf 2}, 031008 (2012)}.
  
\bibitem{interaction1}
W. Witczak-Krempa, M. Knap and D. Abanin, 
{\em Interacting Weyl semimetals: characterization via the topological Hamiltonian and its breakdown}, 
\doi{10.1103/PhysRevLett.113.136402}{Phys. Rev. Lett. 113 (2014) 136402} 
[\arXiv{1406.0843}{cond-mat.str-el}].
 
%\cite{Wang:2012ig}
\bibitem{Wang:2012ig} 
  Z.~Wang and B.~Yan,
{\em Topological Hamiltonian as an Exact Tool for Topological Invariants,}
%  J.\ Phys.\ : Condens.\ Matter {\bf 25}, 155601 (2013)
  \doi{10.1088/0953-8984/25/15/155601}{J.\ Phys.\ : Condens.\ Matter {\bf 25}, 155601 (2013)}
  [\arXiv{1207.7341}{cond-mat.str-el}].
  %%CITATION = doi:10.1088/0953-8984/25/15/155601;%%

\bibitem{burkov}
A.~A.~Burkov, M.~D.~Hook and L.~ Balents,
{\em Topological nodal semimetals}, 
\doi{10.1103/PhysRevB.84.235126}{Phys.\ Rev.\ B {\bf 84}, 235126 (2011)}
[\arXiv{1110.1089}{cond-mat.mes-hall}].
 
 \bibitem{rev1}
C.~Fang, H.~Weng, X.~Dai and Z.~Fang,
{\em Topological nodal line semimetals},
\doi{10.1088/1674-1056/25/11/117106}{Chin. Phys. B 25, 117106 (2016)}
[\arXiv{1609.05414}{cond-mat.mes-hall}].

%\cite{Arutyunov:1998xt}
\bibitem{Arutyunov:1998xt} 
  G.~E.~Arutyunov and S.~A.~Frolov,
  {\em Antisymmetric tensor field on AdS(5),}
 % Phys.\ Lett.\ B {\bf 441}, 173 (1998)
  \doi{10.1016/S0370-2693(98)01136-8}{Phys.\ Lett.\ B {\bf 441}, 173 (1998)}
  [\arXiv{hep-th/9807046}{}].
  
%\cite{Alvares:2011wb}
\bibitem{Alvares:2011wb} 
  R.~Alvares, C.~Hoyos and A.~Karch,
 {\em An improved model of vector mesons in holographic QCD,}
  %Phys.\ Rev.\ D {\bf 84}, 095020 (2011)
  \doi{10.1103/PhysRevD.84.095020}{Phys.\ Rev.\ D {\bf 84}, 095020 (2011)}
  [\arXiv{1108.1191}{hep-ph}].
  %%CITATION = doi:10.1103/PhysRevD.84.095020;%%

\bibitem{Liu:2009dm} 
  H.~Liu, J.~McGreevy and D.~Vegh,
{\em Non-Fermi liquids from holography,}
%Phys.\ Rev.\ D {\bf 83}, 065029 (2011)
 \doi{10.1103/PhysRevD.83.065029}{Phys.\ Rev.\ D {\bf 83}, 065029 (2011)} 
 [\arXiv{0903.2477}{hep-th}].
  %%CITATION = doi:10.1103/PhysRevD.83.065029;%% 
  
 %\cite{Cubrovic:2009ye}
\bibitem{Cubrovic:2009ye} 
  M.~Cubrovic, J.~Zaanen and K.~Schalm,
{\em String Theory, Quantum Phase Transitions and the Emergent Fermi-Liquid,}
 % Science {\bf 325}, 439 (2009)
\doi{10.1126/science.1174962}{Science {\bf 325}, 439 (2009)} 
  [\arXiv{0904.1993}{hep-th}].
  %%CITATION = doi:10.1126/science.1174962;%%

%\cite{Iqbal:2009fd}
\bibitem{Iqbal:2009fd} 
  N.~Iqbal and H.~Liu,
  {\em Real-time response in AdS/CFT with application to spinors,}
  \doi{10.1002/prop.200900057}{ Fortsch.\ Phys.\  {\bf 57}, 367 (2009)}
  [\arXiv{0903.2596}{hep-th}].
  %%CITATION = doi:10.1002/prop.200900057;%%


%\cite{Plantz:2018tqf}
\bibitem{Plantz:2018tqf} 
  N.~W.~M.~Plantz, F.~Garcia Florez and H.~T.~C.~Stoof,
 {\em Massive Dirac fermions from holography,}
%  JHEP {\bf 1804}, 123 (2018)
  \doi{10.1007/JHEP04(2018)123}{JHEP {\bf 1804}, 123 (2018)}
  [\arXiv{1802.04191}{hep-th}].

\bibitem{berry}
M. V. Berry, 
{\em Quantal phase factors accompanying adia- batic changes,}
\href{http://rspa.royalsocietypublishing.org/content/392/1802/45}{Proc. Roy. Soc. Lond. A 392, 45 (1984)}.


\bibitem{Nielsen}  
H. N. Nielsen and M. Ninomiya,
{\em The Adler-Bell-Jackiw anomaly and Weyl fermions in a crystal,} 
\href{https://www.sciencedirect.com/science/article/pii/0370269383915290}{Physics Letters B130 389 (1983)}.


 \end{thebibliography}
%\bibliographystyle{jhepcap}

% %%%%%%%%%%%%%%%%%%%%%%%%%
% %%%%%%%%%%%%%%%%%%%%%%%%%
 
\end{document}